\documentclass[%
reprint,
superscriptaddress,
 amsmath,amssymb,
 aps,
 prx,
]{revtex4-2}

\usepackage{graphicx}
\usepackage{dcolumn}
\usepackage{bm}

\usepackage{xcolor}

\newcommand{\total}{{\rm d}}
\newcommand{\ud}{\mathrm{d}}
\newcommand{\diff}[2]{\frac{\ud {#1}}{\ud {#2}}}
\newcommand{\pdiff}[2]{\frac{\partial #1}{\partial #2}}

\usepackage{booktabs}
\usepackage[newparttoc]{titlesec}
\usepackage{subcaption}
\usepackage[colorlinks=true,allcolors=blue]{hyperref}

\begin{document}

\title{Efficient Quantum Analytic Nuclear Gradients 
with
Double Factorization}

\date{\today}

\author{Edward G. Hohenstein}
\affiliation{
QC Ware Corporation, Palo Alto, CA 94301, USA
}

\author{Oumarou Oumarou}
\affiliation{
Covestro Deutschland AG, Leverkusen 51373, Germany
}

\author{Rachael Al-Saadon}
\affiliation{
QC Ware Corporation, Palo Alto, CA 94301, USA
}

\author{Gian-Luca R. Anselmetti}
\affiliation{
Covestro Deutschland AG, Leverkusen 51373, Germany
}

\author{Maximilian Scheurer}
\affiliation{
Covestro Deutschland AG, Leverkusen 51373, Germany
}

\author{Christian Gogolin}
\email{christian.gogolin@covestro.com}
\affiliation{
Covestro Deutschland AG, Leverkusen 51373, Germany
}

\author{Robert M. Parrish}
\email{rob.parrish@qcware.com}
\affiliation{
QC Ware Corporation, Palo Alto, CA 94301, USA
}

\begin{abstract}
Efficient representations of the Hamiltonian such as double factorization drastically reduce circuit depth or number of repetitions in error corrected and noisy intermediate scale quantum (NISQ) algorithms for chemistry.
We report a Lagrangian-based approach for evaluating relaxed one- and two-particle reduced density matrices from double factorized Hamiltonians, unlocking efficiency improvements in computing the nuclear gradient and related derivative properties.
We demonstrate the accuracy and feasibility of our Lagrangian-based approach to recover all off-diagonal density matrix elements in classically-simulated examples with up to 327 quantum and 18470 total atoms in QM/MM simulations, with modest-sized quantum active spaces. We show this in the context of the variational quantum eigensolver (VQE) in case studies such as transition state optimization, ab initio molecular dynamics simulation and energy minimization of large molecular systems.
\end{abstract}

\maketitle


\section{Introduction}
Analytic derivatives of the energy are essential for efficient evaluation of chemical observables \cite{Bishop1099,Pulay1969,Yamaguchi:1994,Pulay2014}.
Geometry optimization and reaction pathway simulation are routine tasks in computational chemistry which require the repeated evaluation of nuclear derivatives.
Other molecular properties such as the dipole and polarizability can be formulated as derivatives of the energy with respect to electric field perturbations.
Mixed derivatives of nuclear and electric (or magnetic) perturbations define common spectroscopic observables \cite{Jensen:2017}.
While numerical techniques for the nuclear derivatives are often straightforward, they are inefficient and inexact -- the compute time scales linearly with the number of perturbations, which would be three times the total number of atoms in the case of the nuclear gradient.
Only the availability of reasonably well scaling analytic derivatives makes realistic chemical simulations across large length and/or time scales practical and thus they also need to be developed for quantum algorithms for chemistry.

The key to making analytic derivative methods practical is to reduce the complexity of the derivative to be within a multiplicative prefactor of the energy computation \cite{handy1984evaluation}.
For first derivatives, this appears to be possible for all classical electronic structure methods where the derivative is a well-defined quantity.
The challenge in formulating the gradient is in dealing with the nested series of procedures that defines the final energy.
As this pertains to noisy intermediate-scale quantum (NISQ) era quantum computing, the standard implementation of the variational quantum eigensolver (VQE) \cite{Peruzzo:2014:1, McClean:2016:023023, Romero:2018:014008} involves the use of an atom-centered Gaussian orbital basis set, the construction of a set of molecular orbitals, and finally the optimization of the VQE wavefunction.
To evaluate the derivative of the VQE energy, the response of the VQE wavefunction parameters, the molecular orbitals and the underlying basis functions to the perturbation must be considered \cite{rice1986analytic}.
This can be handled elegantly and with a minimal amount of computational effort within a Lagrangian formulation \cite{Helgaker:1982:939}.
A Lagrangian is constructed that contains undetermined multipliers that are used to make the energy stationary with respect to the non-variational wavefunction parameters.
In most electronic structure methods, the linear equations defining the Lagrange multipliers need to be solved only once and an arbitrary number of nuclear or electric field perturbations can then be evaluated.
This removes the dependence of the scaling of the derivative on the number of perturbations, which constitutes the main computational advantage of analytic derivatives over numerical derivatives.

A major challenge for NISQ and FTQC quantum algorithms for chemistry is the complexity of the electron repulsion integral (ERI) contributions to the Hamiltonian, which manifests itself in large circuit depths and/or large numbers of distinct measurement bases and experimental repetitions (shots).
Efficient algorithms for quantum chemistry \cite{PhysRevLett.120.110501, Berry:2019:208, Huggins2021, Motta_2021} work around this problem by means of factorization techniques, which however add additional layers of complexity to the formulation of derivative properties.
The case we consider here is the explicit double factorization (X-DF) of the Hamiltonian \cite{PhysRevLett.120.110501, Berry:2019:208, Huggins2021, Motta_2021, Cohn2021filter}.
This factorization offers several advantages for quantum algorithms in the NISQ era and beyond.
Perhaps most importantly, the diagonal density matrices necessary to compute the expectation value of the energy can be evaluated with a series of simultaneous measurements in each of these bases \cite{Huggins2021}.
By reducing the number of measurements, the number of times the wavefunction must be constructed is immediately reduced.
Additionally, in a finite and limited shot budget setup, shot budgetization is just as important as reducing the number of terms.
X-DF proves to be very reliable in this respect providing a small variance in the distribution of the expectation values \cite{Huggins2021}.

Similar low-rank factorizations are ubiquitous in classical quantum chemistry. The most closely related to X-DF are the density fitting (or resolution of the identity) \cite{Vahtras:1993:514, Feyereisen:1993:359} and Cholesky decomposition approximations \cite{Beebe:1977:683, Roeggen:1986:154, Koch:2003:9481}.
It is the second transformation in X-DF that is omitted in standard algorithms because it is not clear that this would offer any practical advantage on classical hardware.
Other factorizations such as the pseudospectral method \cite{Friesner:1986:1462, Friesner:1987:3522, Friesner:1991:341}, chain-of-spheres \cite{Izsak:2011:144105} and tensor hypercontraction (THC) \cite{Hohenstein:2012:044103, Parrish:2012:224106, Parrish:2013:132505} provide a more flexible decomposition of the ERI tensor, but bear fewer similarities to X-DF.
It should be noted that superficially the THC factorization appears the most similar to X-DF, however, X-DF offers less algebraic flexibility and, thus, cannot reduce the scaling of exchange-like contractions.
Of these integral factorizations, density fitting is probably the most widely applied in chemical simulations.
In large part, this is because potential energy surfaces generated by density-fitted methods are continuous and it is straightforward to develop analytic expressions for their derivatives \cite{Distasio:2007:839, Bostrom:2013:204, Delcey:2015:044110}.
Other methods, such as Cholesky decompositions do not yield continuous potential surfaces in general.
Additional restrictions must be placed on the formation of the Cholesky decomposition to obtain continuous potential surfaces and analytic derivatives thereof \cite{Aquilante:2008:034106, Aquilante:2009:154107, Bostrom:2014:321}.
An advantage of X-DF is that it both defines continuous potential surfaces and its formulation makes it possible to define analytic derivatives.

The focus of our present work is on the development of analytic derivatives of the energy for hybrid quantum classical methods that leverage X-DF Hamiltonians.
There are two particular challenges to the formulation of these derivatives.
First, the factorization and approximation of the ERI tensor introduces new non-variational parameters whose response to perturbations must be considered.
Second, the reduction in the number of measurements to compute the energy afforded by X-DF comes with a catch.
Namely, the off-diagonal elements of the density matrices are not needed to compute the energy, but their effective inclusion is required to evaluate the gradient.
To measure these off-diagonal matrix elements (e.g., by measurement of Pauli strings in the Jordan-Wigner representation) would negate some of the computational advantage offered by X-DF.
Moreover, if X-DF is to be deployed on fermionic quantum simulator hardware, it may be formally impossible to directly measure these off-diagonal matrix elements (e.g., if Pauli gates to directly measure off-diagonal matrix elements in are inadmissible). 
Clearly, an alternative must be developed.
To circumvent this limitation, we take a Lagrangian-based approach to the analytic derivative and introduce undetermined multipliers to account for all of the non-variational parameters that appear in the energy expressions of hybrid quantum classical methods that leverage X-DF Hamiltonians.
Our formulation allows the exact derivative of the energy to be evaluated while retaining the advantages of the X-DF Hamiltonian. Importantly, the Lagrangian formulation allows us to avoid solving for the response of the energy to each perturbation individually.

\section{Theory}
The active space electronic Hamiltonian of a molecular system acting on the fermionic Fock space over some spin adapted orbital basis can be written as
\begin{align}
\hat{H}
=
E_{\rm c}
& +
\sum_{pq}
(p|\hat{h}_{\rm c}|q)
\hat{E}_{pq} \nonumber \\
& +
\frac{1}{2}
\sum_{pqrs}
(pq|rs)
\left(
\hat{E}_{pq}
\hat{E}_{rs}
-
\delta_{qr}
\hat{E}_{ps}
\right),
\end{align}
where $\hat{E}_{pq} = \hat{p}^\dagger \hat{q} + \hat{\bar{p}}^\dagger \hat{\bar{q}}$ are the singlet excitation operators and $\hat{p}^\dagger$ ($\hat{\bar{p}}^\dagger$) is the alpha/spin up (beta/spin down) fermionic creation operator in orbital $p$, with the annihilation operator counterpart $\hat p$.
$E_{\rm c}$ is the frozen core energy, combining the coulomb energy of the nuclei and the frozen core electrons,
$(p|\hat{h}_{\rm c}|q)$ are the active space one-body integrals including contribution from the interaction with the frozen core and $(pq|rs)$ are the two-body integrals.

Double factorization yields a compressed representation \cite{Motta_2021,PhysRevLett.120.110501,Huggins2021,Cohn2021filter} of such Hamiltonian from which energy expectation values for any many-body state $|\Psi\rangle$ can be computed (or estimated from finite shot estimated expectation values) by means of
\begin{align}
\label{eqn:E_xdf_main}
E
\equiv
\langle \Psi | \hat{H} | \Psi \rangle
\equiv
\mathcal{E}
+
\sum_k
\mathcal{F}^\varnothing_{k}
\omega^\varnothing_{k}
+
\sum_{t=1}^{n_t}\sum_{kl}
Z^t_{kl}
\omega^t_{kl},
\end{align}
where we refer to the terms in the last equation as $t$-leaves for $t \in \{\varnothing, 1, \dots, n_t\}$ and call
\begin{align}
\omega^\varnothing_{k} &\equiv \frac{1}{2}
\langle \Psi(\varnothing) |
\hat{Z}_{k}
+
\hat{Z}_{\bar{k}}
| \Psi(\varnothing) \rangle
\\
\omega^t_{kl} &\equiv \frac{1}{8}
\langle \Psi(t) |
\hat{Z}_{k}
\hat{Z}_{l}
-
\hat{I}
\delta_{kl}
+
\hat{Z}_{k}
\hat{Z}_{\bar{l}}
\nonumber\\&\quad+
\hat{Z}_{\bar{k}}
\hat{Z}_{l}
+
\hat{Z}_{\bar{k}}
\hat{Z}_{\bar{l}}
-
\hat{I}
\delta_{\bar{k}\bar{l}}
| \Psi(t) \rangle
\end{align}
the double factorized eigenbasis one- and two-body reduced density matrices and the states $|\Psi(t)\rangle$ are obtained by Jordan Wigner mapping $|\Psi\rangle$ to qubits and rotating the state with a unitary $U^t$ that implements an orthogonal rotation of the spatial orbitals. Furthermore, $\hat{Z}$ is a Pauli-Z operator whose subscript $k$ ($\bar{k}$) indicates that it acts on the qubit corresponding to the alpha (beta) spin orbital corresponding to spatial orbtial $k$.
$\mathcal{E}$ is a constant energy offset containing the core energy $E_c$.

The $U^t$ as well as the vector $\mathcal{F}^\varnothing_{k}$ and the tensor $Z^t_{kl}$ depend on the integrals $(p|\hat{h}_{\rm c}|q)$ and $(pq|rs)$, and in X-DF, they are determined by means of (nested) diagonalizations of these quantities.
The resulting representation is exact for the case $n_t = N (N+1)/2$ where $N$ is the number of active spatial orbitals.
The repesentation can be arbitrarily truncated, which can be meaningful if the $t$-leaves of the factorization are ordered appropriately.
If quantities are to be measured by computing expectation values from finitely many shots a better approach can be to importance-sample the leafs according to their contribution to the overall variance of the estimator.
Compressed double factorization (C-DF) produces representations of the same form but with $U^t$ and $Z^t_{kl}$ determined via a a more elaborate optimization procedure and typically yields a better approximation for the same number of $t$-leaves \cite{Cohn2021filter}.
For more details see supplemental material Section I.A.

The $U^t$ can be efficiently implemented on a quantum computer in linear circuit depth even with only linear nearest neighbor connectivity \cite{THOULESS1960225,PhysRevA.92.062318,PhysRevLett.73.58,PhysRevLett.120.110501} by means of a fabric of Givens rotations $\hat{G}_g$ with angles $\theta^{t}_{g}$ chosen such that
\begin{equation}
\label{eqn:givens_U}
\prod_{g}
\hat G_g(\theta^{t}_{g})
=
U^t,
\end{equation}
and the angles $\theta^{t}_{g}$ can be efficiently computed from $U^t$.

The double factorized eigenbasis one- and two-body reduced density matrices $\omega^\varnothing_{k}$ and $\omega^t_{kl}$ are sufficient to compute the energy but they are insufficient to reconstruct the standard full one and two-body fermionic reduced density matrices (RDMs),
\begin{align}\label{eq:onerdm}
\gamma_{pq}
\equiv
\frac{\mathrm{d} E}{\mathrm{d} ( p | \hat h_c | q )}
=
\langle
\Psi
|
\hat{E}_{pq}
|
\Psi
\rangle 
\end{align}
\begin{align}\label{eq:twordm}
\Gamma_{pqrs}
\equiv 
\frac{\mathrm{d} E}{\mathrm{d} (pq|rs)}
=
\frac{1}{2}
\langle
\Psi
|
\hat{E}_{pq}
\hat{E}_{rs}
-
\delta_{qr}
\hat{E}_{ps}
|
\Psi
\rangle.
\end{align}

The first advantage of using equation \eqref{eqn:E_xdf_main} to compute or estimate the energy is that the number of distinct bases in which measurements need to be performed is bounded by $n_t+1$ where $n_t\leq{n(n+1)}/{2}$.
The circuit depth only increases linearly to accommodate the necessary transformation to the orbital basis. 
At the same time X-DF reduces variance and thereby shows improved error rates with greatly reduced  shot counts \cite{Huggins2021}.

Key to the efficient evaluation of the analytic gradient is avoiding the explicit computation of perturbation dependent response. One approach to obtaining the derivative of the energy is to simply apply the chain rule and evaluate all the partial derivatives that appear. While this will certainly lead to a correct result, it also leads to an amount of unnecessary work that scales linearly with the number of perturbations. For example, one would need to solve for $\frac{\partial\/V^t_{pq}}{\partial\xi}$ and $\frac{\partial\/U^t_{pk}}{\partial\xi}$ in order to account for the derivatives of the implicit parameters introduced in the X-DF two-electron operator. This is a well-known problem in electronic structure theory and can be avoided by using a Lagrangian formulation of the derivative \cite{Helgaker:1982:939}.

These issues are avoided by use of the Lagrangian formalism.
The idea behind the Lagrangian approach to analytic derivatives is to construct a Lagrangian whose value is equal to the energy (the quantity we wish to differentiate). To that, an undetermined multiplier is added corresponding to every non-variational, implicitly-defined quantity that contributes to the energy expression. These multiply a constraint chosen from the implicit definition of the parameters.
We can then solve a linear system to make the Lagrangian stationary with respect to variation in the implicitly-defined parameters. 

The X-DF Lagrangian takes the form
\begin{align}
\label{eqn:lagrangian}
\mathcal{L}_{\mathrm{X\mbox{-}DF}} &\equiv \mathcal{E}
+
\sum_k
\mathcal{F}^\varnothing_{kk}
\omega^\varnothing_{k}
+
\sum_{t}^{T}\sum_{kl}
Z^t_{kl}
\omega^t_{kl}
\nonumber \\
&
+
\sum_{p > k}
\eta_{pk}^{\varnothing}
\left[
\left(
\prod_{g}
\hat{G}_g
( \theta_{g}^{\varnothing} )
\right)_{pk}
-
U_{pk}^{\varnothing}
\right]
\nonumber \\
&
+
\sum_{t}^{T}
\sum_{p > k}
\eta_{pk}^{t}
\left[
\left(
\prod_{g}
\hat{G}_g
( \theta_{g}^{t} )
\right)_{pk}
-
U_{pk}^{t}
\right]
\nonumber \\
&
+
\sum_{k > k'}
\mu_{kk'}^{\varnothing}
\mathcal{F}_{kk'}^{\varnothing}
\nonumber \\
&
+
\sum_{t}^{T}
\sum_{k > k'}
\mu_{kk'}^{t}
\lambda_{kk'}^{t}
+
\sum_{t > t'}
\nu^{tt'}
g^{tt'},
\end{align}
where $\mathcal{F}^\varnothing_{kk'} = \sum_{pq} U^\varnothing_{pk} \mathcal{F}^\varnothing_{pq} U^\varnothing_{qk'}$ with
\begin{equation}
\mathcal{F}^\varnothing_{pq} 
=
(p|\hat{h}_{\rm c}|q)
+
\sum_r
(pq|rr)
-
\frac{1}{2}
\sum_r
(pr|qr) ,
\end{equation}
and $Z^t_{kl}$, $\lambda^t_{kk'}$, and $g^{tt'}$ are defined in a similar way in terms of nested orthogonal transformations of $(pq|rs)$. $\eta^t_{pk}$, $\mu^t_{pk}$, and $\nu^{tt'}$ are Lagrange multipliers.
The Lagrangian is designed such that it is stationary with respect to the $U^t_{pk}$, $\theta^t_g$, $\lambda^t_{kk'}$, and $g^{tt'}$ resulting from the X-DF procedure and the construction of the Givens fabrics and for those $U^t_{pk}$, e.g., the $\mathcal{F}^\varnothing_{kk'}$ are diagonal.

The $\eta^t_{pk}$ multipliers ensure that the Givens fabrics actually implement the corresponding orbital rotation $U^t$.
The $\mu^t_{pk}$ make the Lagrangian stationary with respect to variations in the $U^t_{pk}$.
The $\nu^{tt'}$ multipliers depend on derivatives of the two-body contribution to the energy as well as the $\mu^t_{pk}$ multipliers.
These multipliers couple the $t$-leaves together and, in the case of truncated X-DF they account for rotations between $t$-leaves included in the definition of the Hamiltonian and those that were excluded. These multipliers make the Lagrangian stationary with respect to variation in the eigenvectors of the two-electron integrals

The full Lagrangian of any orbital based active space method will contain further contributions from orbital response, but those will depend on details of and can be taken care of on the level of the method that was used to construct the orbitals.
A complete description of each term appearing in the Lagrangian can be found in the Supporting Information in addition to a detailed derivation of the equations that must be solved to reach stationarity in all of the nonvariational parameters.

There is an interesting interplay between the $\eta$ multipliers used to make the Lagrangian stationary with respect to variations in the parameterization of the Givens operators and the $\mu$ and $\nu$ multipliers corresponding to the X-DF matrix elements.
The quantities that appear in the $\eta$ constraint terms carry no explicit dependence on the perturbation, therefore they have no direct contribution to the analytic derivative. However, the $\eta$ multipliers form the right-hand side of the linear system that defines the $\mu$ multipliers.
In the case of $\mu^\varnothing$, this term does carry explicit dependence on the perturbation through the one-body Hamiltonian matrix elements.
In the case of $\mu^t$, it again carries no explicit dependence on the perturbation, but forms a portion of the right-hand side of the linear system defining $\nu$.
The $\nu$ terms carries explicit dependence on the perturbation through the ERIs. The nested $\eta^\varnothing$ and $\mu^\varnothing$ multipliers recover the off-diagonal one-body density matrix elements, while the series of $\eta^t$, $\mu^t$ and $\nu$ allow the full two-body density matrix to be recovered.
In this way, we are able to recover the information content of the full density matrices through a Lagrangian-based approach without ever needing to measure the full density matrices directly.

We demonstrate how this can be done for the one body density matrix $\gamma_{pq}$ and leave further details to the SI.
We will define the one-body density matrix as the derivative of the Lagrangian (which has been made stationary with respect to all the X-DF non-variational parameters: $\theta_g^\varnothing$, $\theta_g^t$, $U^\varnothing_{pk}$, $U^t_{pk}$, $V^t_{pq}$) with respect to the frozen core operator, $(p|h_c|q)$. We denote this Lagrangian (which omits the $z_{pq}$ and $x_{pq}$ multipliers) as $\mathcal{L}_\mathrm{X\mbox{-}DF}$.
Then with the Kronecker delta $\delta_{pq}$
\begin{align}\label{eq:mu_llprime}
\pdiff{\mathcal{L}_\mathrm{X\mbox{-}DF}}{(p|h_c|q)} & =
\nonumber \\
\gamma_{pq}
&
= \delta_{pq}
+ \sum_k U^\varnothing_{pk} \omega_k^\varnothing U^\varnothing_{qk}
+ \sum_{kk'} U^\varnothing_{pk} \mu_{kk'}^\varnothing U^\varnothing_{qk'}.
\end{align}
The $\omega_k^\varnothing$ were already measured to determine the energy, the $U^\varnothing_{pk}$ are a result of the factorization, and it remains to determine the $\mu_{kk'}$.
To this end, one starts from $\diff{\mathcal{L}_{\mathrm{X\mbox{-}DF}}}{U^\varnothing_{pk}} = 0$ and arrives at
\begin{align}
\mu_{ll'}^{\varnothing}
=
\frac{
\eta_{ll'}^{\varnothing}
-
\eta_{l'l}^{\varnothing}
}{\mathcal{F}_{l'l'}^{\varnothing} - \mathcal{F}_{ll}^{\varnothing}}.
\label{eqn:mu_main}
\end{align}
The $\eta_{ll'}^{\varnothing}$ multipliers are in turn defined via the linear system of equations
\begin{equation}
\sum_{p > k}
\eta^\varnothing_{pk}
\underbrace{
\diff{\mathcal{O}}{\theta_{g'}}
\left [
\prod_{g}
\hat G_{g} (\theta^\varnothing_{g})
\right ]_{pk}
}_{A^\varnothing_{g',pk}}
=
-
\diff{E}{\theta^t_{g'}},
\label{eqn:eta_main}
\end{equation}
which can be solved by inverting the square $N(N-1)/2$ matrix $-A^\varnothing_{g',pk}$ and contracting the result against $\diff{E}{\theta^t_{g'}}$, which is just the derivative of the energy with respect to the Givens angles and can be measured by means of the parameter shift rule.

An important consideration is the stability of this procedure and the potential presence of singularities in the linear equations defining the Lagrange multipliers.
Singularities in Lagrange multipliers defined by \eqref{eqn:mu_main} (or similar equations) can appear if eigenvalues of the underlying operator are degenerate due to symmetry.
In this case, the numerator of the Lagrange multiplier should be zero by symmetry as well and, therefore, these blocks of the Lagrange multipliers should be zero by L'H\^{o}pital's rule.
Removal of these singularities can be accomplished through adaptation of the Lagrangian for molecular symmetry.
The matrix $A^\varnothing_{g,pk}$ (and related matrices) may also be singular.
To avoid these issues, we solve the linear system in equation \eqref{eqn:eta_main} by pseudoinversion of this matrix; negligible eigenvalues are set to zero.
In the practical application of the analytic derivatives that follows, we did not encounter any numerical issues related to singularities in the linear equations defining the Lagrange multipliers.
Should instabilities issues arise they can be worked around in the following way: the energy (Equation \eqref{eqn:E_xdf_main}) is a linear function of $(p|\hat h_c|q)$ and $(pq|rs)$, so it does not matter at what position the derivatives in \eqref{eq:onerdm} and \eqref{eq:twordm} are evaluated.
That is, the reduced density matrices are a function of the state only and do not depend on $(p|\hat h_c|q)$ or $(pq|rs)$.
For the purpose of computation of the reduced density matrices one is free to substitute other tensors that minimize the variance of the final quantity one is interested in computing (i.e., by contraction with the density matrices).
For example, Equation \eqref{eqn:mu_main} suggests to choose a substitute for $(p|\hat h_c|q)$ with in equally spaced eigenvalues (these eigenvalues replace $\mathcal{F}^\varnothing_{ll'}$) and an eigenbasis that makes $A^\varnothing_{g',pk}$ well conditioned.
Investigating to which extent this allows to further reduce the influence of shot noise will be left to a future investigation.

\section{Computational Details}
We have implemented the analytic derivatives of VQE energies with X-DF Hamiltonians within an in-house C++/Python quantum circuit simulator code. Hamiltonian matrix elements, molecular orbitals and molecular orbital response are handled by the {\tt Lightspeed} + {\tt TeraChem} \cite{Seritan:2021:e1494} code stack.
Molecular mechanics computations are performed with {\tt OpenMM} \cite{OpenMM} interfaced through {\tt Lightspeed}. VQE wave functions were constructed with quantum-number-preserving (QNP) gate fabrics \cite{Anselmetti2021} to create VQE wavefunctions similar to $k$-UpCCGSD \cite{OGorman:2019, Lee:2019:311}. Molecular orbitals are obtained with Hartree-Fock (HF) and fractional occupation number Hartree-Fock (FON-HF) \cite{Gidopoulos:1993:1067, Gidopoulos:2002:328}. Further details on the systems we simulate are provided as needed in the remainder of the manuscript and Supporting Information.

\section{Results}
To validate the correctness of our formulation and implementation of the derivatives of the X-DF Hamiltonian, we compare the analytic derivatives to numerical derivatives obtained from finite difference computations. For this test, we use the X-DF Hamiltonian to construct VQE wavefunctions for butadiene with a 4 electron in 4 orbital (4\textit{e}, 4\textit{o}) active space and a 6-31G** basis set. There are four cases we will consider to determine the correctness of the analytic derivative. First, there is the case where the X-DF is exact (i.e., the eigenbasis of the two-electron operator is not truncated) and the VQE parameterization is sufficiently flexible to recover the exact ground state energy. In the second case, the X-DF of the Hamiltonian is approximate, but the VQE is still converged to the exact ground state of that approximate Hamiltonian. In the third case, the X-DF is exact, but the VQE solution is approximate. Finally, we consider the case where both the X-DF and the VQE wavefunctions are approximate. Our Lagrangian formulation of the analytic derivative accounts for truncation of the $t$-leaves thus accounting for both exact and approximate X-DF Hamiltonians. The analytic derivatives of exact and approximate VQE energies do not require additional effort due to the variational nature of the method.

The complete results of the comparison between our analytic and numerical derivatives can be found in the Supporting Information. In all of the cases described above, we obtain agreement within 10$^{-6}$ $E_{\rm h}/a_0$ indicating the correctness of our formulation and implementation of the analytic derivatives. Importantly, we ``stress test'' our response equations by using a severe truncation of the X-DF Hamiltonian. In the two cases where the X-DF Hamiltonian is approximate, we neglect all eigenvalues below 10$^{-2}$ $E_{\rm h}$ resulting in only 4 of 10 $t$-leaves being retained. The agreement between the analytic and numerical derivatives is comparable with both the truncated and untruncated X-DF Hamiltonians, indicating that our formulation of the analytic derivative correctly handles these additional response contributions.
While these numerical tests give us confidence in the correctness of derivation and implementation, they do not speak directly to the robustness of the approach, i.e., how stable the implementation is across larger regions of the potential energy surface.

One important application of analytic derivatives is in force evaluation during {\em ab initio} molecular dynamics (AIMD) simulations. These simulations also provide a numerical check for the consistency of the energy and its derivative. In order to perform conservative dynamics simulations within the microcanonical ensemble (i.e., the $NVE$ ensemble), the forces must be obtained from the exact derivative of the potential: $F = - \frac{{\rm d}E}{{\rm d}R}$. If this is the case, it is possible to perform energy-conserving AIMD simulations.

\begin{figure}
    \centering
    \includegraphics[width=0.5\textwidth]{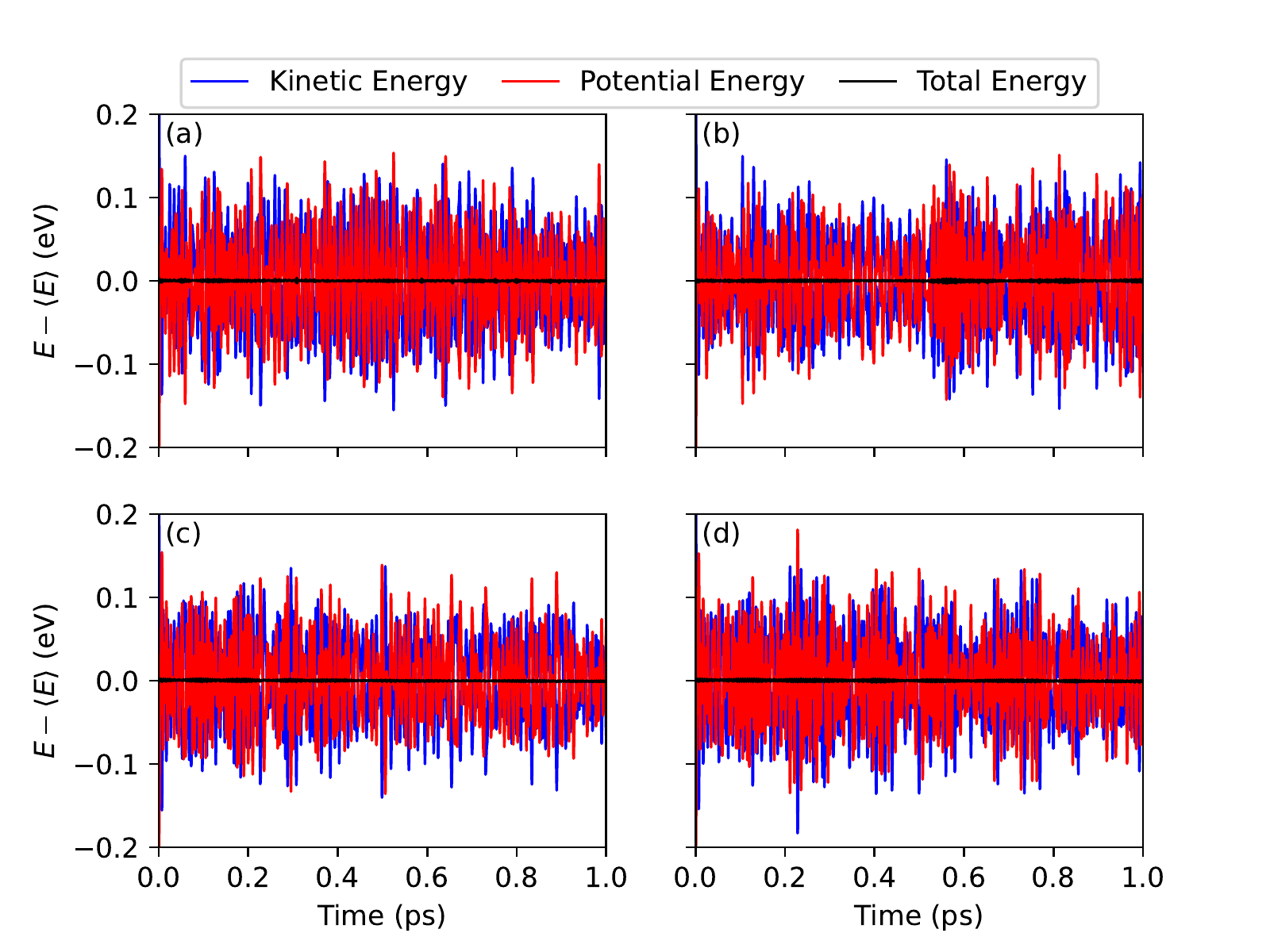}
    \caption{{\em Ab initio} molecular dynamics simulation of butadiene using a (4\textit{e}, 4\textit{o})  active space with VQE energies obtained using an X-DF Hamiltonian and a 6-31G$^{\ast\ast}$ basis set. Integration of the classical equations of motion was performed using velocity Verlet with a 0.5 fs time step. (a) The VQE energy was converged to the exact ground state energy and the X-DF Hamiltonian was not truncated. (b) The VQE energy was converged to the exact ground state energy and the eigenvalues of the ERI tensor were truncated at 10$^{-1}$ au. (c) The VQE energy was not converged to the exact ground state energy and the X-DF Hamiltonian was not truncated. (d) The VQE energy was not converged to the exact ground state energy and the eigenvalues of the ERI tensor were truncated at 10$^{-3}$ au.
    }
    \label{fig:aimd}
\end{figure}

We again use butadiene as a test of our analytic derivatives (using the same level of theory as described previously). This time, we perform a 1 ps AIMD simulation using velocity Verlet integration \cite{Swope:1982:637} with a 0.5 fs time step. In this case, we use the fluctuations (and possible drift) of the total energy as a check for the correctness of the analytic derivatives of the energy. In addition, this tests the stability of the implementation across the regions of the potential energy surface explored by the AIMD trajectory. Discontinuities on the potential surface or instabilities in the response equations will lead to a loss of energy conservation. As before, we test the four cases of exactness and approximation in X-DF and VQE. The results of these simulations are shown in Figure \ref{fig:aimd}. In each case, the AIMD simulations conserve energy indicating that the response equations needed to evaluate the analytic derivatives remain stable throughout. One should note that when the X-DF Hamiltonian is approximate, the potential energy surface only remains continuous if the two-electron operator eigenvalues at the truncation boundary do not become degenerate. That is, the ``diabatic'' character of the $t$-leaves must not change discontinuously with geometry. These eigenvalues define surfaces (in analogy to the Born-Oppenheimer potential energy surface) and those eigensurfaces may have conical intersections between them. In the event that such intersections occur, it must be the case that either both intersecting eigenvalues are retained or omitted in the X-DF Hamiltonian to maintain continuity of the potential energy surface.

\begin{figure}
    \centering
    \includegraphics[width=0.5\textwidth]{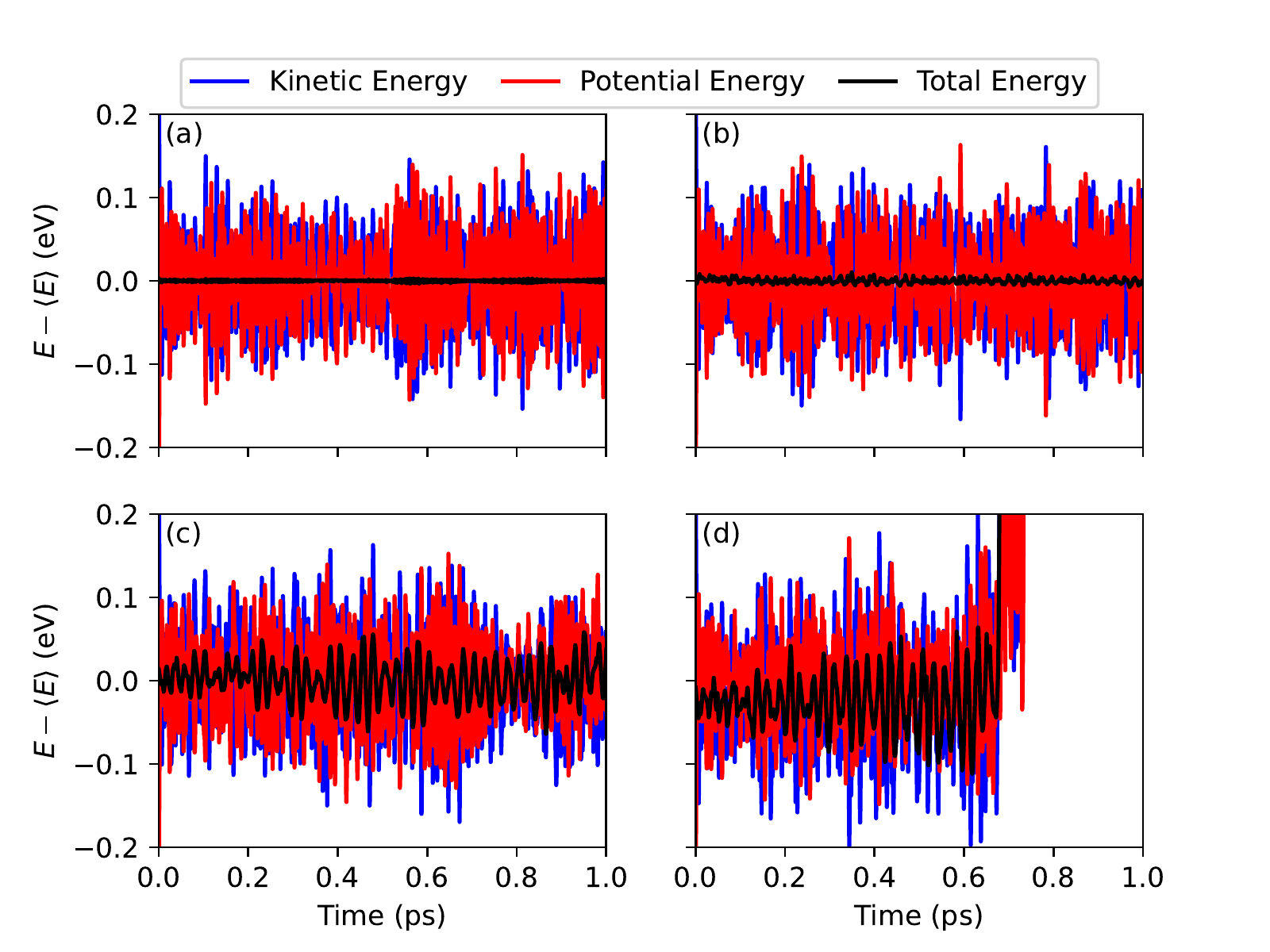}
    \caption{{\em Ab initio} molecular dynamics simulation of butadiene using a (4\textit{e}, 4\textit{o}) active space with VQE energies obtained using an X-DF Hamiltonian and a 6-31G$^{\ast\ast}$ basis set. Integration of the classical equations of motion was performed using velocity Verlet with a 0.5 fs time step. In all cases, the eigenvalues of the ERI tensor were truncated at 10$^{-1}$ au in the X-DF Hamiltonian and the VQE energy was converged to the exact ground state energy. (a) No approximations were made to the analytic gradient expressions, this is equivalent to Figure \ref{fig:aimd}(b). (b) The $\eta^\varnothing$ and $\mu^\varnothing$ multipliers were set to zero during the evaluation of the analytic gradient. (c) The $\eta^t$ and $\mu^t$ multipliers were set to zero during the evaluation of the analytic gradient. (d) The $\nu^t$ multipliers were set to zero during the evaluation of the analytic gradient.
    }
    \label{fig:aimd_broke}
\end{figure}

To test the importance of the response contributions to the analytic derivative, we intentionally introduce errors in to the derivative expression by artificially setting some of the Lagrange multipliers to zero. Using these approximate derivatives, we perform AIMD simulations as before and test the energy conservation. The results of these simulations are shown in Figure \ref{fig:aimd_broke}. The response of the one-electron terms in the X-DF Hamiltonian can be removed by setting either the $\eta^\varnothing$ or the $\mu^\varnothing$ multipliers to zero. Remarkably, the neglect of these response contributions to the gradient does not have an outsized impact on energy conservation (although the AIMD simulation with these errors introduced clearly does not conserve energy as well as those using the exact analytic derivatives). What is remarkable about this is that setting these multipliers to zero effectively neglects the off-diagonal matrix elements of the one-body density matrix; one might have expected this to have a more deleterious effect on the dynamics of the system. In the third case, the $\eta^t$ or, equivalently, the $\mu^t$ multipliers are set to zero. This removes significant contributions to the two-body density matrix and greatly affects the energy conservation. In this case, fluctuations in the total energy are nearly as large as the fluctuations in kinetic and potential energy. Despite this, the butadiene molecule remains intact throughout the 1 ps simulation. In the final case, the $\nu^{tt^\prime}$ multipliers are set to zero (this also removes contributions from $\eta^t$ and $\mu^t$). With these errrors introduced to the derivative, it is no longer possible to perform the AIMD simulation for the full 1 ps. This indicates the importance of the response contributions to the analytic derivative -- particularly the responses originating in the factorization of the two-electron operator.

\begin{figure}
    \centering
    \includegraphics[width=0.5\textwidth]{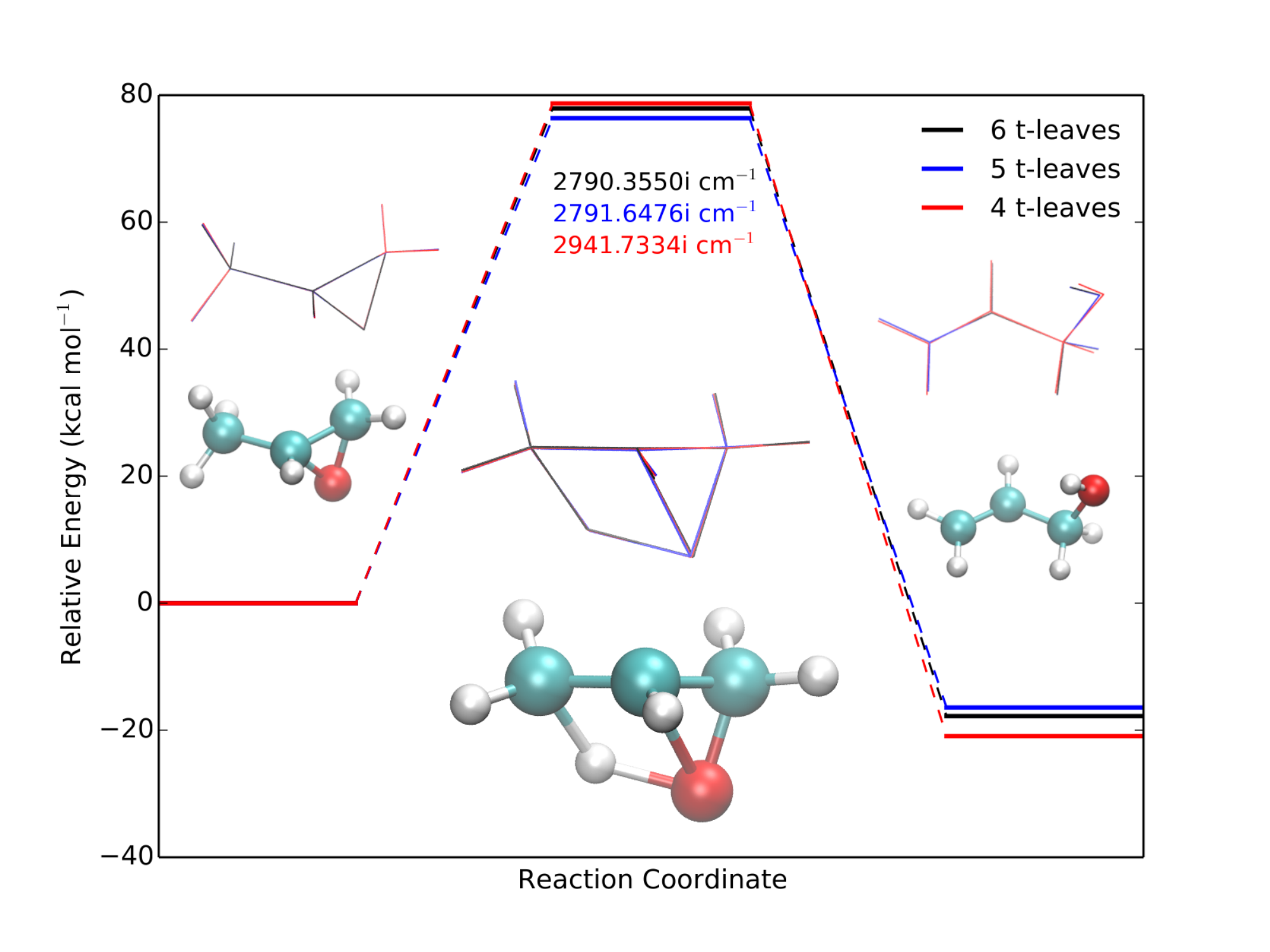}
    \caption{Optimization of reactant, product and transition state geometries for the propylene oxide to allyl alcohol isomerization reaction. The potential energy surface is obtained from VQE using a (4\textit{e}, 3\textit{o}) active space using an X-DF Hamiltonian and a 6-31G$^{\ast\ast}$ basis set. The full X-DF Hamiltonian contains 6 $t$-leaves; results are presented with truncated X-DF expansions as well. The relative energies of these geometries are given in kcal mol$^{-1}$. At the transition state, the frequency of the imaginary mode (i.e., the negative Hessian eigenvalue) is given in cm$^{-1}$. The optimized structures are shown overlaid to compare the accuracy of the geometries obtained from different truncations of the X-DF Hamiltonian.}
        \label{fig:ts_optimization}
\end{figure}

Another important application of analytic derivatives of the energy is in the location and characterization of stationary points on the potential energy surface. As a representative example of this, we examine the isomerization reaction of propylene oxide to allyl alcohol (Figure \ref{fig:ts_optimization}) \cite{Dubnikova:2000:4489}. The potential energy surface is obtained by VQE with a (4\textit{e}, 3\textit{o}) active space and a 6-31G$^{\ast\ast}$ basis set -- the VQE parameterization is sufficient to obtain the exact ground state energy. At this level of theory, we optimize the geometries of the reactant (propylene oxide), product (allyl alcohol) and the transition state connecting them. The optimizations were converged tightly with the maximum absolute value of any component of the derivative being less than 10$^{-6}$ $E_{\rm h}/a_0$. To characterize the transition state, the nuclear Hessian was evaluated by finite differences of the analytic derivatives; from that Hessian, the harmonic vibrational frequencies could be determined. These optimizations were performed using exact and approximate X-DF Hamiltonians. Since the active space is relatively small, the exact X-DF Hamiltonian contains only 6 $t$-leaves. As a result, truncation of this expansion has a significant impact on the potential energy surface. We were only able to obtain optimized geometries when at least 4 $t$-leaves were retained in the factorization.

It has been well established in other methods leveraging approximate factorizations of the two-electron operator that relative energies can be obtained more accurately than absolute energies. This is particularly important to applications in chemistry where the absolute energy of a molecule is rarely -- if ever -- of practical importance. Case in point, the reaction energy and reaction barrier we obtain for the propylene oxide to allyl alcohol isomerization are relative energies that are chemically relevant. We find that the reaction barrier is slightly underestimated (by 1.5 kcal mol$^{-1}$) when 5 $t$-leaves are retained and slightly overestimated (by 0.8 kcal mol$^{-1}$) when 4 $t$-leaves are retained. Similar results are obtained for the reaction energy. The magnitude of the reaction energy is underestimated by 1.4
kcal mol$^{-1}$ when 5 $t$-leaves are retained and overestimated by 3.1 kcal mol$^{-1}$ when 4 $t$-leaves are retained. The imaginary frequency of the transition state is well reproduced with 5 $t$-leaves; an error on the order of 1 cm$^{-1}$ is introduced in that case. With 4 $t$-leaves, the error increases to 150 cm$^{-1}$. In all cases, the geometries of these stationary points are in good agreement. This indicates that not only can X-DF be used in the characterization of chemical reactions, but also that truncated X-DF can be applied.

\begin{figure}
    \centering
    \includegraphics[width=0.5\textwidth]{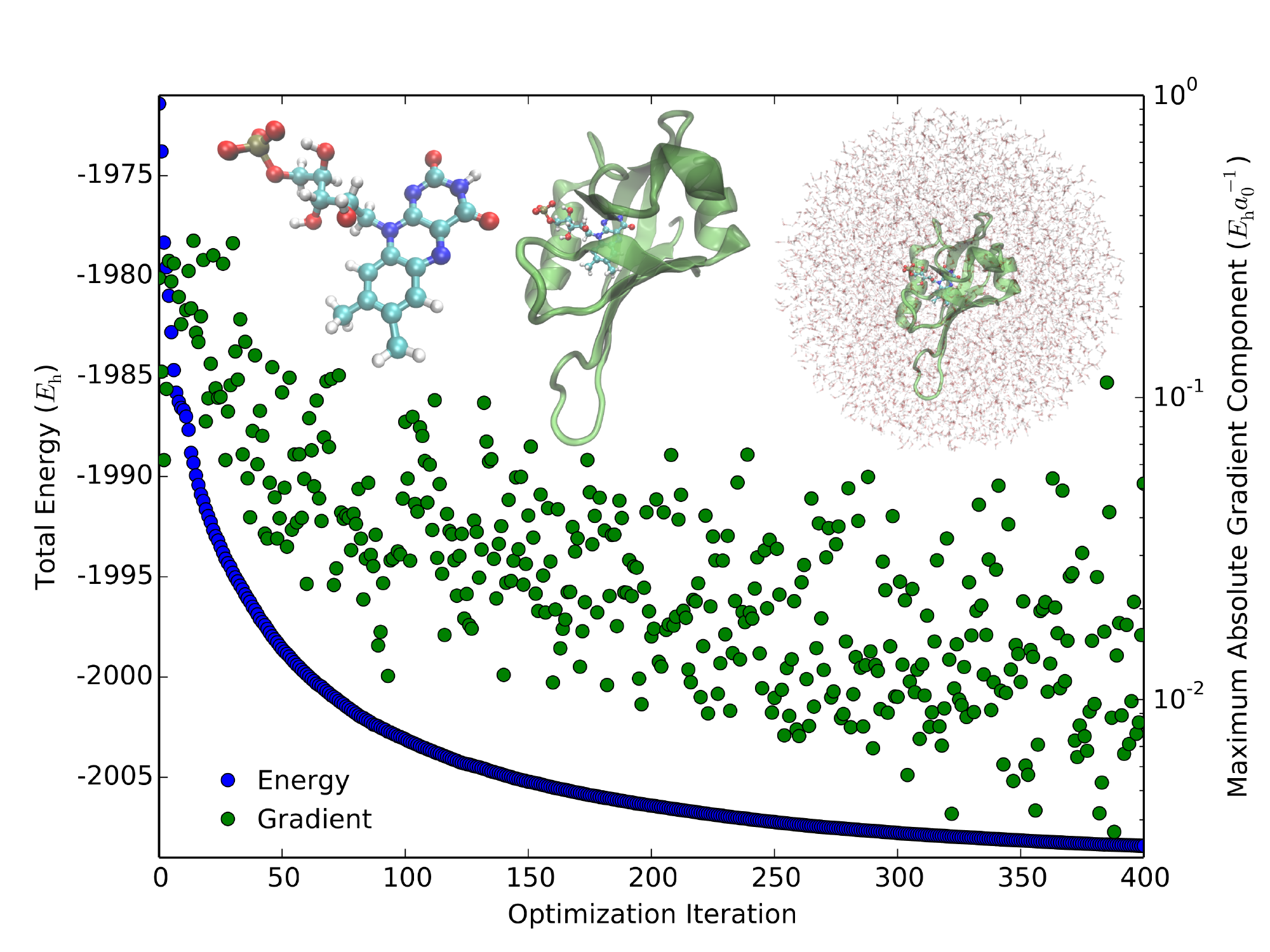}
    \caption{QM/MM optimization of a flavin-binding light, oxygen or voltage (LOV) domain from the phototropin module  of {\em Adiantum} phy3 in its dark state \cite{Crosson:2001:2995}. The QM region contains the flavin mononucleotide chromophore (50 atoms), while the remaining atoms are treated at the MM level. This chromophore is noncovalently bound to the protein. Including solvent molecules, this system contain 18470 atoms. The QM region is described by VQE using a (4\textit{e}, 4\textit{o}) active space, an X-DF Hamiltonian without truncation and a 6-31G basis set; molecular orbitals were obtained from HF computation. The total energy of the system as well as the maximum absolute component of the gradient are reported for each iteration of an L-BFGS optimization performed in Cartesian coordinates.}
    \label{fig:lov_optimization}
\end{figure}

For an implementation of analytic gradients to be practical, it should be able to be applied to molecular systems with many nuclear degrees of freedom without a significant increase in the computational cost relative to the underlying energy evaluation. An example where this is particularly important is in hybrid quantum mechanics molecular mechanics (QM/MM) computations \cite{Warshel1976}. Here, the number of total nuclear degrees of freedom may be many orders of magnitude larger than what is contained in the quantum mechanical region alone. A representative example is shown in Figure \ref{fig:lov_optimization}. Here a QM region containing 50 atoms is coupled to an MM region containing 18470 atoms. The electronic Hamiltonian of the QM system contains the point charges associated with the nuclei in the QM region as well as all of the MM atoms. In the context of analytic derivatives, this means that there are 55410 nuclear perturbations that must be considered. A formulation of the gradient requiring explicit evaluation of perturbation-dependent response would be intractable. Our Lagrangian-based approach, however, allows QM/MM computations to be performed at the expense of evaluating the additional one-electron integrals (and their derivatives) and their subsequent contraction with the appropriate density matrices and Lagrange multipliers.

To demonstrate the ability to treat extended molecular systems, we optimized the geometry of a flavin mononucleotide-containing protein \cite{Crosson:2001:2995} in the presence of explicit solvent (Figure \ref{fig:lov_optimization}). We report the first 400 steps of an limited-memory Broyden–Fletcher–Goldfarb–Shanno (L-BFGS) optimization performed in Cartesian coordinates. The chromophore (containing 50 atoms) comprised the QM region and all 55410 nuclear degrees of freedom were relaxed during the optimization. Consistent progress was made in minimizing both the energy and gradient. This indicates the suitability of our approach for simulations of large molecular systems. 

\begin{figure}
    \centering
    \includegraphics[width=0.5\textwidth]{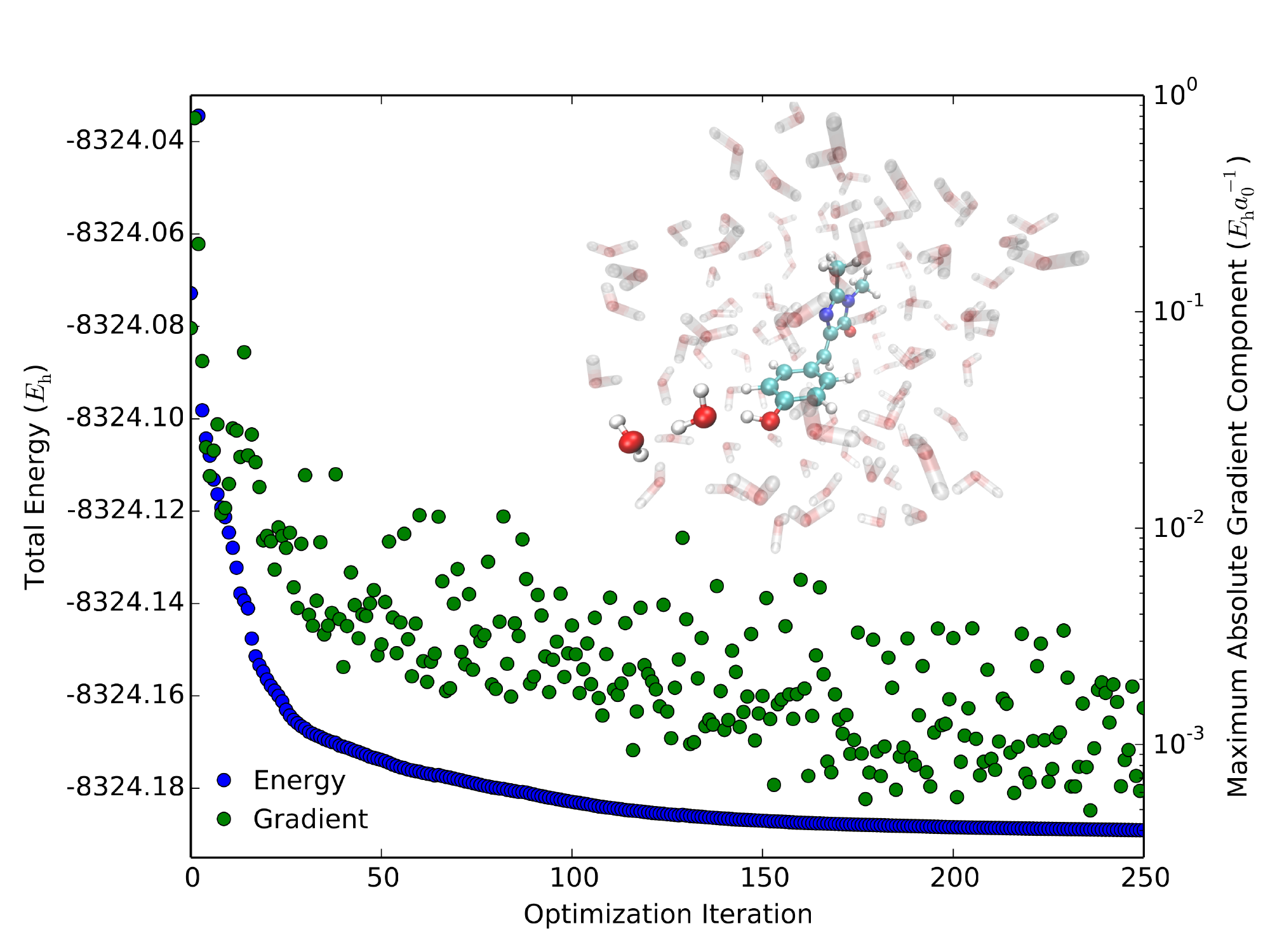}
    \caption{Optimization of the anionic HBDI chromophore solvated by 100 water molecules including an extra proton. The entire is system is treated with VQE using a (4\textit{e}, 3\textit{o}) active space, an X-DF Hamiltonian without truncation and a 6-31G$^{\ast\ast}$ basis set; orbitals were generated with FON-HF using Gaussian broadening of the orbital energy levels and temperature parameter of 0.2 a.u. The optimization is started from a minimum energy conical intersection between the ground and first excited state of the chromophore. The total energy of the system as well as the maximum absolute component of the gradient are reported for each iteration of an L-BFGS optimization performed in Cartesian coordinates.}
    \label{fig:hbdi_optimization}
\end{figure}

Another important consideration for implementations of analytic gradients is whether or not they scale appropriately with the size of the molecular system. It has been established that the matrix elements for active space methods such as VQE can be generated using the same algorithms used for Fock matrix construction in Hartree-Fock (HF) or Kohn-Sham density functional theory (KS-DFT) \cite{Hohenstein:2015:224103}. This means that -- provided the size of the active space is constant -- the VQE method should scale like HF or KS-DFT as the system size is increased. This also applies to analytic derivatives of active space methods; the analytic derivative can be cast in terms of derivatives of Fock-like quantities \cite{Hohenstein:2015:224103, Hohenstein:2015:014111}. All of this is to say that an implementation of VQE gradients should be able to be performed on large molecular systems. An example of such a system is shown in Figure \ref{fig:hbdi_optimization}. Here, the HBDI (4-hydroxybenzylidene-1,2-dimethylimidazolinone) chromophore of the green fluorescent protein (GFP) is solvated by 100 water molecules \cite{Mandal:2004:1102}. The starting structures were obtained from minimum energy conical intersection optimization using floating occupation molecular orbital complete active space configuration interaction (FOMO-CASCI) \cite{Hohenstein:2016:174110}. Since these optimizations were possible with classical electronic structure methods, it is essential that they are also possible using VQE with an X-DF Hamiltonian.

The optimizations shown in Figure \ref{fig:hbdi_optimization} highlight a few important aspects of our analytic gradient formulation. First, and perhaps most obvious, is that it was possible to perform many optimization cycles for a molecular system containing over 300 atoms and 2800 basis functions. This is because our implementation scales appropriately with the size of the overall system as well as the size of the active space. This optimization intentionally used a starting structure near a conical intersection to emphasize the ability of methods like VQE for describing electronic structure in the presence of strong correlation. The optimization proceeded smoothly despite starting at such a structure. Finally, the starting structure we chose was for an anionic HBDI chromophore that had previously dissociated a proton (bonded to two neighboring water molecules). Over the course of the first 50 optimization cycles, the proton transfers back to the HBDI chromophore that is now relaxing on its ground electronic state. By using explicit solvent molecules treated quantum mechanically, it is possible to observe such proton transfer reactions during optimization or molecular dynamics simulations. During the final 200 optimization cycles, the HBDI chromophore returns from the twisted geometry that minimizes the energy of the conical intersection to a planar geometry that minimizes the energy of the ground state.

\section{Conclusions}

We have presented a framework for evaluating analytic derivatives from X-DF Hamiltonians.
The increased efficiency in quantum computation offered by X-DF is preserved in our derivative formulation.
One of the principle advantages of X-DF is that is reduces the number of measurements required by evaluating the energy with density matrices expressed in the bases ($t$-leaves) determined by X-DF.
While the information contained in these density matrices is sufficient to determine the energy, it is not sufficient to determine derivatives of the energy. Our Lagrangian formulation of the derivative recovers the missing information without the need to explicitly measure the missing off-diagonal contributions to the density matrices (and thereby limiting the computational advantage offered by X-DF).
Importantly, our formulation of the analytic derivative does not require perturbation-dependent response, i.e., the quantum computing effort scales with the number of orbitals treated on the quantum computer only and not with the total number of atoms.
This is an essential characteristic of a method that is intended to have practical applications in chemistry.

Our initial tests of the analytic derivatives used VQE wavefunctions, however, our Lagrangian framework can be extended to other quantum algorithms.
For example, it would be straightforward to extend this to derivatives of multistate contracted VQE (MC-VQE) energies \cite{Parrish:2019:230401} where the VQE parameters are not variationally optimized for a single state.
One would simply need to include additional Lagrange multipliers to make the state-specific MC-VQE energies stationary with respect to their wavefunction parameters \cite{Parrish2021mcvqegradient}.
Further, this general approach can be extended to other factorizations of the Hamiltonian such as the compressed double factorization (C-DF) \cite{Cohn2021filter}.
Finally, our work also opens the door to exploiting the advantages of compressed representations of the Hamiltonian when computing quantities beyond energy in an error corrected setting.
This is a necessary step towards making quantum computation more broadly applicable in chemical simulation.

\textbf{Acknowledgements:}
We thank Fotios Gkritsis for valuable discussions.
QC Ware Corp. acknowledges generous funding from Covestro for the undertaking of this project.
Covestro acknowledges funding from the
German Ministry for Education and Research (BMBF)
under the funding program quantum technologies as part
of project HFAK (13N15630).

\textbf{Conflict of Interest:} A portion of the quantum Lagrangian gradient methodology pertains to a provisional patent application filed jointly by Covestro and QC Ware Corp.
EGH and RMP own stock/options in QC Ware Corp. 


%

\cleardoublepage

\titleformat{\part}
  {\normalfont}
  {}
  {8pt}
  {\Large\bfseries}
\makeatletter
\@addtoreset{section}{part}
\makeatother
\part{Supporting Information}

\section{
\label{sec:problem}
Problem Statement}

In this section, we describe the scope of the work presented in this paper. We focus on the case of analytic nuclear gradients of variational energies determined from the explicit double factorization of the Hamiltonian but it will be apparent how the methods outlined here can be used to compute full relaxed active space one and two body reduced density matrices that can serve as input parameters for the computation of other quantities in the same way as is can be done in other active space methods.

\subsection{
\label{sec:xdf}
Hamiltonian Factorization}

The electronic Hamiltonian can be normal-ordered with respect to
some closed-shell Slater determinant representing the frozen core
electrons and written as:
\begin{align}
\hat{H}
=
E_{\rm c} 
& +
\sum_{pq}
(p|\hat{h}_{\rm c}|q)
\hat{E}_{pq} \nonumber \\
& +
\frac{1}{2}
\sum_{pqrs}
(pq|rs)
\left(
\hat{E}_{pq}
\hat{E}_{rs}
-
\delta_{qr}
\hat{E}_{ps}
\right),
\end{align}
where $\hat{E}_{pq}$ are the usual singlet excitation operators,
$E_{\rm c}$ is the energy of the frozen core determinant,
and $(p|\hat{h}_{\rm c}|q)$ are the matrix elements of the so-called 
frozen core operator,
\begin{equation}
(p|\hat{h}_{\rm c}|q) 
= 
(p|\hat{h}|q)
+
\sum_i^{\rm core}
\left[
2 (pq|ii)
-
(pi|qi)
\right].
\end{equation}
Explicit double factorization of the Hamiltonian yields an exact or approximate and compressed  
representation of such Hamiltonian by means of diagonalization of the integral tensors.
First, consider the modified
one-electron integrals
\begin{equation}
(p|\hat{\kappa}|q)
=
(p|\hat{h}_{\rm c}|q)
-
\frac{1}{2}
\sum_r
(pr|qr).
\end{equation}
These can be trivially eigendecomposed as
\begin{equation}
(p|\hat{\kappa}|q)
=
\sum_k 
U^\varnothing_{pk}
f^\varnothing_{k}
U^\varnothing_{qk}.
\end{equation}
The two-electron integrals can be factorized first by
eigendecomposing the integral tensor
\begin{equation}
\label{eqn:eri_eigen}
(pq|rs)
=
\sum_t
V^t_{pq}
g^t
V^t_{rs}.
\end{equation}
Subsequently, the eigenvectors can be decomposed for each
``eigenleaf'' (term in the above sum), which will be referred to as a $t$-leaf throughout
the remainder of this work,
\begin{equation}
V^t_{pq}
=
\sum_k
U^t_{pk}
\lambda^t_k
U^t_{qk}.
\end{equation}
Combining these nested decompositions, we arrive at the double
factorized form of the two-electron integrals.
\begin{equation}
\label{eqn:df}
(pq|rs)
=
\sum_{tkl}
U^t_{pk}
U^t_{qk}
\underbrace{
\lambda^t_k
g^t
\lambda^t_l}
_{Z^t_{kl}}
U^t_{rl}
U^t_{sl}
\end{equation}
The approach outlined above defines the explicit double 
factorization (X-DF) method. Other approaches of obtaining
double factorized two-electron integral \eqref{eqn:df} have
been developed, but we will concentrate on X-DF in the present work.
X-DF can be either exact or approximate;
this depends on whether or not all the $t$-leaves are retained or those
with small eigenvalues \eqref{eqn:eri_eigen} have been removed.
Now, we can write the Hamiltonian in terms of these factorized 
one- and two-electron integrals as
\begin{equation}
\label{eqn:H_old}
\hat{H}
=
E_{\rm c} 
+
\sum_{k}
f^\varnothing_{k}
\hat{E}_{kk}(\varnothing) +
\frac{1}{2}
\sum_{tkl}
Z^t_{kl}
\hat{E}_{kk}(t)
\hat{E}_{ll}(t).
\end{equation}
Notice that the dependence of $\hat{E}_{kk}(t)$ on the $t$-leaf
specifies that these operators are applied in a rotated orbital
basis defined by the eigendecompositions shown above.

In the present work, we make a few additional modifications to the
Hamiltonian. If we consider the form of the $\hat{E}_{kk}$ operators
within the Jordan-Wigner mapping,
\begin{equation}
\hat{E}_{kk}
=
\hat{I}
-
\frac{1}{2}
\left(
\hat{Z}_{k}
+
\hat{Z}_{\bar{k}}
\right),
\end{equation}
then we can define new zero- and one-body terms that include effective
contributions from the one- and two-body terms in \eqref{eqn:H_old},
\begin{equation}
\mathcal{E} 
= 
E_{\rm c} 
+ 
\sum_p 
(p|\hat{h}_{\rm c}|p)
+
\frac{1}{2}
\sum_{pq}
(pp|qq)
-
\frac{1}{4}
\sum_{pq}
(pq|pq),
\end{equation}
\begin{equation}
\mathcal{F}_{pq} 
=
(p|\hat{h}_{\rm c}|q)
+
\sum_r
(pq|rr)
-
\frac{1}{2}
\sum_r
(pr|qr).
\end{equation}
As before, the effective one-electron operator can be eigendecomposed as
\begin{equation}
\mathcal{F}^\varnothing_{pq}
=
\sum_k 
U^\varnothing_{pk}
\mathcal{F}^\varnothing_{k}
U^\varnothing_{qk}.
\end{equation}
Now, the Hamiltonian can be written as
\begin{align}
\label{eqn:H_xdf}
\hat{H}
&
=
\mathcal{E}
-
\frac{1}{2}
\sum_k
\mathcal{F}^\varnothing_{k}
\left(
\hat{Z}_{k}
+
\hat{Z}_{\bar{k}}
\right)
+
\frac{1}{8}
\sum_{tkl}
Z^t_{kl}
\nonumber \\
&
\times
\left(
\hat{Z}_{k}
\hat{Z}_{l}
-
\hat{I}
\delta_{kl}
+
\hat{Z}_{k}
\hat{Z}_{\bar{l}}
+
\hat{Z}_{\bar{k}}
\hat{Z}_{l}
+
\hat{Z}_{\bar{k}}
\hat{Z}_{\bar{l}}
-
\hat{I}
\delta_{\bar{k}\bar{l}}
\right).
\end{align}
Here, we suppress the dependence of the $Z_k$ operators on the
orbital basis of the particular $t$-leaf on which they are acting in
the presentation of the equation, but it remains as in \eqref{eqn:H_old}. Throughout this work, we will use the Hamiltonian in
the form presented in \eqref{eqn:H_xdf}.

\subsection{
\label{sec:ci}
Electronic Wavefunction}

In the present work, we will consider only wavefunctions that 
variationally minimize the energy such as those obtained from
the variational quantum eigensolver (VQE). In this approach,
the wavefunction is defined by a parameterized unitary operator acting
on some reference state
\begin{equation}
|\Psi_{\rm VQE}\rangle 
= 
\hat{U}(\vec{\phi})|\Psi_{0}\rangle.
\end{equation}
As the name of the method implies, the parameters, $\vec{\phi}$ are obtained by variationally minimizing the expectation value of the
energy
\begin{equation}
E 
= 
\underset{\vec{\phi}}{\rm min} \;
\langle \Psi_{0} | 
\hat{U}(\vec{\phi})^\dagger \hat{H} \hat{U}(\vec{\phi})
| \Psi_{0}\rangle.
\end{equation}
The key to our subsequent development of analytic derivatives of the
energy is that the VQE energy is stationary with respect to its
parameters,
\begin{equation}
\frac{\partial\/E}{\partial\vec{\phi}} = 0.
\end{equation}
Therefore, we do not need to consider the response of the VQE 
wavefunction to the perturbation under consideration.
We note that -- although we do not consider such cases in the present work
-- the analytic derivatives we formulate could apply to other, non-variational methods provided that Lagrange multipliers can be added
to make the energy stationary with respect to changes in its parameters, as was done in \cite{Parrish2021mcvqegradient} for the case of multi state contracted VQE.

\subsection{
\label{sec:vqe}
X-DF/VQE}

The use of the double factorized Hamiltonian imparts a particular structure on the VQE density matrices. First, consider the usual molecular orbitals,
\begin{equation}
\phi_p({\bf r}) = \sum_\mu C_{\mu\/p} \phi_\mu({\bf r}).
\end{equation}
Here, some some set of atomic orbitals ($\{|\phi_\mu\rangle\}$) are transformed by the molecular orbital coefficient matrix, $C_{\mu\/p}$. The
double factorized Hamiltonian introduces additional orbital bases -- one for each $t$-leaf,
\begin{equation}
\phi_k^t({\bf r}) = \sum_p U^t_{pk} \phi_p({\bf r}).
\end{equation}
The application of the Hamiltonian requires a series of transformations between these new orbital bases. These transformations are performed by a parameterized ``Givens'' operator, $\hat{G}(\{\theta_g^t\})$. Any unitary matrix (of rank $N$) can be represented as a product of $N(N-1)/2$ Givens rotations. 
This can be done in a number of different ways, resulting in different circuit layouts and depths.
This fact is used to determine the set of angles, $\{\theta_g^t\}$, that define each Givens operator. The angles are 
chosen such that there is a unique Givens operator for each $t$-leaf,
\begin{equation}
\label{eqn:givens_U_supp}
\left [
\prod_{g}
\hat G_g (\theta^{t}_{g})
\right ]^{t}_{pk}
=
U^t_{pk}.
\end{equation}
The classical matrix elements of the X-DF Hamiltonian can be obtained from transformations using the $U^t_{pk}$ matrices, while transformations of
the state vector on the quantum computer require the Givens operator to
be applied as a series of Givens circuits. Now we can write the Hamiltonian in its complete form,
\begin{align}
\label{eqn:H_xdf_G}
\hat{H}
=
\mathcal{E}
&
-
\frac{1}{2}
\sum_k
\mathcal{F}^\varnothing_{k}
\hat G^\dagger (\theta^{\varnothing}_{g})
\left(
\hat{Z}_{k}
+
\hat{Z}_{\bar{k}}
\right)
\hat G (\theta^{\varnothing}_{g})
\nonumber \\
&
+
\frac{1}{8}
\sum_{tkl}
Z^t_{kl}
\times
\hat G^\dagger (\theta^{t}_{g})
\big(
\hat{Z}_{k}
\hat{Z}_{l}
-
\hat{I}
\delta_{kl}
+
\hat{Z}_{k}
\hat{Z}_{\bar{l}}
\nonumber \\
&
+
\hat{Z}_{\bar{k}}
\hat{Z}_{l}
+
\hat{Z}_{\bar{k}}
\hat{Z}_{\bar{l}}
-
\hat{I}
\delta_{\bar{k}\bar{l}}
\big)
\hat G (\theta^{t}_{g}),
\end{align}
where the transformations to the orbital eigenbases of each $t$-leaf have been made explicit.

In methods based on the usual electronic Hamiltonian, the energy can be obtained from the one- and two-body density matrices in the molecular orbital basis,
\begin{align}
E
=
E_{\rm c} 
+
\sum_{pq}
(p|\hat{h}_{\rm c}|q)
\gamma_{pq}
+
\sum_{pqrs}
(pq|rs)
\Gamma_{pqrs}.
\end{align}
Here, the molecular orbital-based density matrices have their usual 
definition,
\begin{align}
\gamma_{pq}
=
\langle
\Psi
|
\hat{E}_{pq}
|
\Psi
\rangle
\end{align}
\begin{align}
\Gamma_{pqrs}
=
\frac{1}{2}
\langle
\Psi
|
\hat{E}_{pq}
\hat{E}_{rs}
-
\delta_{qr}
\hat{E}_{ps}
|
\Psi
\rangle.
\end{align}
While it is certainly possible to form the molecular orbital density matrices from the X-DF Hamiltonian, it is more advantageous to work in the natural eigenbasis representation of each $t$-leaf,
\begin{align}
\label{eqn:E_xdf}
E
=
\mathcal{E}
+
\sum_k
\mathcal{F}^\varnothing_{k}
\omega^\varnothing_{k}
+
\sum_{tkl}
Z^t_{kl}
\omega^t_{kl}.
\end{align}
Here, the density matrices are defined as
\begin{align}
\label{eqn:OPDM_xdf}
\omega^\varnothing_{k}
=
-
\frac{1}{2}
\langle
\Psi
|
\hat G^\dagger (\{\theta^{\varnothing}_{g}\})
\left(
\hat{Z}_{k}
+
\hat{Z}_{\bar{k}}
\right)
\hat G (\{\theta^{\varnothing}_{g}\})
|
\Psi
\rangle,
\end{align}
and
\begin{align}
\label{eqn:TPDM_xdf}
\omega^t_{kl}
=
&
\frac{1}{8}
\langle
\Psi
|
\hat G^\dagger (\{\theta^{t}_{g}\})
\big(
\hat{Z}_{k}
\hat{Z}_{l}
-
\hat{I}
\delta_{kl}
+
\hat{Z}_{k}
\hat{Z}_{\bar{l}}
\nonumber \\
+
&
\hat{Z}_{\bar{k}}
\hat{Z}_{l}
+
\hat{Z}_{\bar{k}}
\hat{Z}_{\bar{l}}
-
\hat{I}
\delta_{\bar{k}\bar{l}}
\big)
\hat G (\{\theta^{t}_{g}\})
|
\Psi
\rangle.
\end{align}
In the remainder of this work, we will work exclusively in terms of these eigenbasis density matrices.

\subsection{
\label{sec:gradients}
Nuclear Gradients}

The main focus of this work is the evaluation of analytic derivatives of the energy with respect to the Cartesian positions of the nuclei. The derivation is specialized for the case of atom-centered Gaussian basis functions. This means that the position of the basis functions
is tied to the position of the nuclei. As a result, derivatives of integrals over operators with
no dependence on the nuclei (electronic kinetic or overlap operators, for example) will still have dependence on the nuclear positions. Derivatives of the overlap integrals in the atomic orbital basis provide a representative example,
\begin{equation}
\frac{{\rm d}}{{\rm d}R}
(\mu|\nu)
=
(\frac{{\rm d}}{{\rm d}R}\mu|\nu)
+
(\mu|\frac{{\rm d}}{{\rm d}R}\nu).
\end{equation}
As a consequence of these derivatives surviving, straightforward application of the Hellmann-Feynman theorem is impossible within Gaussian orbital formulations of electronic
structure theory unless a complete (i.e., infinite) basis is applied.

\subsection{
\label{sec:fon_orbitals}
Fractional Occupation Number Hartree-Fock}

In this work, we obtain the molecular orbitals that underly the X-DF/VQE energies from fractional occupation number Hartree-Fock (FON-HF) \cite{Gidopoulos:1993:1067, Gidopoulos:2002:328}, which includes standard HF as a special case.
This method introduces temperature dependent fractional occupations of the orbitals. Commonly, a Fermi-Dirac distribution is used to obtain
the fractional occupations,
\begin{equation}
n_i
=
\frac{2}{1+{\rm exp}[(\epsilon_i-\epsilon_f)/kT]},
\end{equation}
where $\epsilon_i$ and $n_i$ are the orbital energy and fraction occupation of orbital $i$, respectively. The Fermi energy, $\epsilon_f$, is chosen such that the fractional occupations conserve the total number of electrons,
\begin{equation}
N = \sum_i n_i.
\end{equation}
It is also possible to obtain the fractional occupations from a Gaussian broadening of the orbital energy levels or other distributions.

The motivation for using FON-HF orbitals is that they often provide a good approximation to
variationally optimized orbitals while being nearly identical in computational cost to Hartree-Fock (HF) orbitals. Canonical HF orbitals are a poor choice for strongly correlated systems, because they suffer from bias towards the single determinant for which they are optimized. In extreme cases, HF orbitals can artificially break degeneracies that exist in the system (for example, exact degeneracies between the highest occupied and lowest unoccupied orbital cannot be described). FON-HF, by contrast, does not suffer from these issues; in the case of exact degeneracies, FON-HF will simply place equal fractional occupations in each of the degenerate orbitals. A final note is that although we focus on the choice of FON-HF orbitals in this work, the analytic derivative expressions we provide can be trivially extended to most other definitions of molecular orbitals, provided they are fully decoupled from the VQE energy.

\section{Inapplicability of the Hellmann-Feynman Theorem}

At first glance, analytic derivatives of the energy appear trivial to evaluate in light of the Hellmann-Feynman theorem, which states that the derivative of the energy is simply the expectation value of the derivative Hamiltonian,
\begin{equation}
\frac{{\rm d}E}{{\rm d}R}
=
\langle
\Psi
|
\frac{{\rm d}\hat{H}}{{\rm d}R}
|
\Psi
\rangle.
\end{equation}
However, upon closer examination, it becomes clear that the Hellmann-Feynman theorem rarely has practical relevance. First, the Hellmann-Feynman derivative does not include all contributions
from atom-centered Gaussian functions, in particular, the derivatives of the basis function overlap that are necessary to maintain orthonormality in the molecular orbitals. Second, the
Hellmann-Feynman theorem assumes a fully variational wavefunction (i.e., one in which the energy is minimized with respect to all parameters). In the present work, we consider fully variational VQE ans\"{a}tze, but do not minimize the VQE energy with respect to the molecular orbitals. As a result it is essential to include the response of the molecular orbitals to the perturbation in
the analytic derivative. Here, we obtain the molecular orbitals from FON-HF, which minimize the Mermin free energy -- not the VQE energy. As a result, orbital response contributions to the analytic gradient beyond the Hellmann-Feynman theorem must be considered. We should note that these considerations are common to many electronic structure methods and are in no way unique to X-DF/VQE.

A more interesting point is that the Hellmann-Feynman theorem does not apply for X-DF/VQE in places where it usually would be exact. Take, for example, a case where the VQE ansatz covers all available variational free parameters. Consider also a perturbation that does not change the underlying Gaussian basis functions -- perhaps an electric field perturbation. In that case, the Hellmann-Feynman theorem would hold and the derivative of the energy is equivalent to the one-particle density matrix traced against the one-body electric field derivative integrals (i.e., dipole integrals). Here, for a perturbation in the $x$ direction,
\begin{equation}
\frac{{\rm d}E}{{\rm d}\varepsilon_x}
=
\sum_{pq}
\gamma_{pq}
(p|\hat{\mu}_x|q).
\end{equation}
However, in the X-DF/VQE case, we do not assume that the full density matrices are available, only the eigenbasis density matrices. As a result, application of the Hellmann-Feynman in the case analogous to what is shown above is no longer exact,
\begin{equation}
\frac{{\rm d}E}{{\rm d}\varepsilon_x}
\neq
\frac{{\rm d}\mathcal{E}}{{\rm d}\varepsilon_x}
+
\sum_k
\frac{{\rm d}\mathcal{F}^\varnothing_{k}}{{\rm d}\varepsilon_x}
\omega^\varnothing_{k}.
\end{equation}
The loss of exactness can be traced to the representation of the density matrix in the eigenbasis. This representation of the density is a projection of the full density to the
minimal amount required to recover the energy. This representation of the density alone is
insufficient in general for obtaining properties other than the energy.

To illustrate how the projection of the density affects property computations, consider
a general example,
\begin{equation}
o
=
\sum_{pq}
A_{pq}
B_{pq}.
\end{equation}
Now, if we eigendecompose matrix ${\bf A}$,
\begin{equation}
\label{eqn:eig_A}
A_{pq} = \sum_k U_{pk} a_k U_{qk},
\end{equation}
we can write the original summation in the eigenbasis of ${\bf A}$ as
\begin{equation}
o
=
\sum_{k}
a_{k}
\underbrace{
\left[
\sum_{pq}
U_{pk}
B_{pq}
U_{qk}
\right]
}
_{B_{kk}}.
\end{equation}
Notice that we now have a dot product between the eigenvalues of ${\bf A}$ and the 
diagonal elements of ${\bf B}$ after transformation to the eigenbasis of ${\bf A}$.
Note that unless $[{\bf A},{\bf B}]=0$, ${\bf B}$ is not diagonal in the eigenbasis
of ${\bf A}$. Now, if we transform back to the original basis using \eqref{eqn:eig_A}
and $\tilde{B}_{pq} = \sum_{k} U_{pk} B_{kk} U_{qk}$, we find
\begin{equation}
o
=
\sum_{pq}
A_{pq}
\tilde{B}_{pq}.
\end{equation}
In this case, ${\bf B} \neq {\bf \tilde{B}}$ and we have shown that only a projection of the matrix ${\bf B}$ is required to maintain equality. If we consider a second matrix, ${\bf C}$, where $[{\bf A},{\bf C}] \neq 0$, following the same logic we find that
\begin{equation}
\sum_{pq}
C_{pq}
B_{pq}
\neq
\sum_{pq}
C_{pq}
\tilde{B}_{pq}.
\end{equation}
This is because the projection of ${\bf B}$ is lossy.

What we have shown in general above is what we must contend with when we work in terms of the eigenbasis density matrices (i.e., \eqref{eqn:E_xdf}).
We only retained enough information in the density matrices to compute the energy. The remainder of the information that is needed must be obtained by other means and it turns out that taking derivatives with respect to the parameters of the Givens operators is a suitable approach. Another way of stating this problem is that by writing the energy in terms of only the diagonal elements of the density matrix, we have lost some of the usual orbital invariances.

\section{Lagrangian Formulation of the Gradient}

We must consider the response of several non-variational parameters introduced in the X-DF method in order to evaluate the analytic derivative of the energy. These parameters are the $U^\varnothing_{pk}$, $U^t_{pk}$ and $V^t_{pq}$ tensors that define the X-DF matrix elements. In addition, the X-DF/VQE derivative also depends on $\theta_g^\varnothing$ and $\theta_g^t$, which parameterize the Givens operators used to transform to the eigenbasis of each $t$-leaf. Finally, the derivative of the energy also depends on the usual quantities such as the underlying molecular orbitals and the one- and two-electron integrals in the atomic orbital basis.

Treating the response of these non-variational parameters to the perturbation in an efficient manner is essential to the development of a formulation of the analytic derivative that can be applied to practical chemical problems. From the chain rule, it appears that the response of all of these free parameters to the perturbation is required to evaluate the derivative of the energy,
\begin{align}
\frac{\total\/E}{\total\/R}
=
\frac{\partial\/E}{\partial\/R}
& 
+
\frac{\partial\/E}{\partial\/U^\varnothing_{pk}}
\frac{\total\/U^\varnothing_{pk}}{\total\/R}
+
\frac{\partial\/E}{\partial\/U^t_{pk}}
\frac{\total\/U^t_{pk}}{\total\/R}
+
\frac{\partial\/E}{\partial\/V^t_{pq}}
\frac{\total\/V^t_{pq}}{\total\/R}
\nonumber
\\
&
+
\frac{\partial\/E}{\partial\/\theta^\varnothing_{g}}
\frac{\total\/\theta^\varnothing_{g}}{\total\/R}
+
\frac{\partial\/E}{\partial\/\theta^t_{g}}
\frac{\total\/\theta^t_{g}}{\total\/R}
+
\frac{\partial\/E}{\partial\/C_{\mu\/p}}
\frac{\total\/C_{\mu\/p}}{\total\/R}.
\end{align}
While direct evaluation of this equation will produce the desired result, the computational cost will scale with the number of perturbations. There are several approaches to simplifying such equations to remove the explicit perturbation dependence that go by several names. Perhaps the most systematic approach to removing perturbation dependence is to construct a Lagrangian. The value of the Lagrangian is equal to the energy (the quantity we wish to differentiate) and it contains undetermined multipliers which allow partial derivatives with respect to the non-variational parameters to be set to zero. That is, we enforce
\begin{align}
\mathcal{L} = E,\;\;\;
\frac{\partial\mathcal{L}}{\partial\/U^\varnothing_{pk}}&=0,\;\;\;
\frac{\partial\mathcal{L}}{\partial\/U^t_{pk}}=0,\;\;\;
\frac{\partial\mathcal{L}}{\partial\/V^t_{pq}}=0,\;\;\;
\nonumber \\
\frac{\partial\mathcal{L}}{\partial\/\theta^\varnothing_{g}}&=0,\;\;\;
\frac{\partial\mathcal{L}}{\partial\/\theta^t_{g}}=0,\;\;\;
\frac{\partial\mathcal{L}}{\partial\/C_{\mu\/p}}=0.\;\;\;
\end{align}
Such a Lagrangian can be constructed by adding constraints that are satisfied by the definition of each of these non-variational quantities. The complete Lagrangian we will use to define the
analytic gradient is
\begin{align}
\label{eqn:lagrangian_supp}
\mathcal{L} & = \mathcal{E}
+
\sum_k
\mathcal{F}^\varnothing_{k}
\omega^\varnothing_{k}
+
\sum_{t}^{T}\sum_{kl}
Z^t_{kl}
\omega^t_{kl} 
\nonumber \\
&
+
\sum_{p > k}
\eta_{pk}^{\varnothing}
\left[
\left(
\prod_{g}
\hat{G}_g
( \theta_{g}^{\varnothing} )
\right)_{pk}
-
U_{pk}^{\varnothing}
\right]
\nonumber \\
&
+
\sum_{t}^{T}
\sum_{p > k}
\eta_{pk}^{t}
\left[
\left(
\prod_{g}
\hat{G}_g
( \theta_{g}^{t} )
\right)_{pk}
-
U_{pk}^{t}
\right]
\nonumber \\
&
+
\sum_{k > k'}
\mu_{kk'}^{\varnothing}
\mathcal{F}_{kk'}^{\varnothing}
+
\sum_{t}^{T}
\sum_{k > k'}
\mu_{kk'}^{t}
\lambda_{kk'}^{t}
+
\sum_{t > t'}
\nu^{tt'}
g^{tt'}
\nonumber \\
&
+
\sum_{p > q}
z_{pq}
f_{pq}
+
\sum_{p \leq q}
x_{pq}
[\delta_{pq} - (p|q) ].
\end{align}
Here, we have introduced seven sets of undetermined multipliers. The $\eta$
multipliers use the definition of the Givens operators \eqref{eqn:givens_U_supp} to constrain the values of
$\{\theta\}$. The $\mu$ multipliers use the representation of off-diagonal elements of $\mathcal{F}^{\varnothing}$
and $\lambda^t$ in their eigenbases to constrain $U^\varnothing$ and $U^t$ to be the
eigenvectors of these matrices. We use the $\nu$ multipliers to constrain $V$ to be the eigenvectors of the two-electron integrals, by using the off-diagonal elements of the 
two-electron integrals in its eigenbasis as a constraint. Here, we use $T$ to denote where
the factorization of the two-electron integrals has been truncated. The final two multipliers,
$z$ and $x$, constrain the convergence condition and orthonormality of the underlying molecular orbitals. Here, our constrain is that the Fock matrix, $f_{pq}$, is diagonal and that the overlap of the molecular orbitals, $(p|q)$, is an identity matrix. These constraints and equations defining the undetermined multipliers will be discussed in later sections of this document.

\subsection{Constraints on the Givens Operator}

Let us say we have an antisymmetric matrix of size $N$,
\begin{equation}
X_{rs}
=
-
X_{sr}
\
\forall
\
s,\ r.
\end{equation}
Through the exponential map, this generates a special orthogonal matrix, 
\begin{equation}
O_{pq}
\equiv
\left [
\exp( \hat X )
\right ]_{pq}
\
\in
\
\mathcal{SO}(N).
\end{equation}
Through a Givens decomposition, we can define the special orthogonal
matrix $U_{pk}$.
The pivot structure of the Givens matrices, e.g., rectangle \cite{clements2016optimal},
diamond \cite{PhysRevLett.120.110501}, or otherwise, can be chosen freely and is considered to be fixed,
\begin{equation}
O_{pq} [\{ \theta_{g} \}]
\equiv
\left [
\prod_{g}
\hat G_{g} (\theta_{g})
\right ]_{pq}
\
\in
\
\mathcal{SO}(N)
\end{equation}
Note that here $\hat G_{g} (\theta_{g} )$ is the one-body image of the $g$-th
Givens operator.

We are able to use this to construct Givens decompositions that represent $U^\varnothing_{pk}$ and $U^t_{pk}$.
\begin{equation}
\left [
\prod_{g}
\hat G_g (\theta^{\varnothing}_{g})
\right ]^\varnothing_{pk}
=
U^\varnothing_{pk}
\end{equation}
\begin{equation}
\left [
\prod_{g}
\hat G_g (\theta^{t}_{g})
\right ]^t_{pk}
=
U^t_{pk}
\end{equation}
These equalities form the basis of the constraints that enter the X-DF/VQE
Lagrangian \eqref{eqn:lagrangian_supp}. To solve for the undetermined $\eta$ multipliers, we set the partial derivatives of the Lagrangian with respect to $\theta$ to zero. That is, the equations for $\eta^\varnothing_{pk}$ are defined by ${\partial\mathcal{L}}/{\partial\/\theta^\varnothing_{g}}=0$
and the equations for
$\eta^t_{pk}$ are defined by ${\partial\mathcal{L}}/{\partial\/\theta^t_{g}}=0$.

In the Lagrangian, only the energy \eqref{eqn:E_xdf} and the constraint terms themselves carry explicit dependence on $\theta^\varnothing_{g}$ and $\theta^t_{g}$. Therefore, solving for the $\eta^\varnothing_{pk}$ and $\eta^t_{pk}$ undetermined multipliers is the first step in making the Lagrangian stationary. Equations for the Lagrange multipliers are determined by,
\begin{equation}
\label{eqn:eta0}
\sum_{p > k}
\eta^\varnothing_{pk}
\underbrace{
\diff{\mathcal{O}}{\theta_{g'}}
\left [
\prod_{g}
\hat G_{g} (\theta^\varnothing_{g})
\right ]_{pk}
}_{A^\varnothing_{g',pk}}
=
-
\diff{E}{\theta^\varnothing_{g'}}
\end{equation}
and
\begin{equation}
\sum_{p > k}
\eta^t_{pk}
\underbrace{
\diff{\mathcal{O}}{\theta_{g'}}
\left [
\prod_{g}
\hat G_{g} (\theta^t_{g})
\right ]_{pk}
}_{A^t_{g',pk}}
=
-
\diff{E}{\theta^t_{g'}}
\end{equation}
The matrices $A^\varnothing_{g,pk}$ and $A^t_{g,pk}$ are square of dimension $N (N - 1) / 2$ and can be inverted to obtain $\eta^\varnothing_{pk}$ and $\eta^t_{pk}$. Note that we treat each $t$-leaf separately, thus treating $A^t_{g,pk}$ as a stack of matrices rather than a third-order tensor. In practice, we find that these equations can be accurately solved by pseudoinverting the ${\bf A}$ matrices as they may become singular in cases with high symmetry depending on orbital ordering and the pivot structure of the givens decomposition.

\subsection{Constraints on the X-DF Eigenbases}

In the X-DF/VQE Lagrangian \eqref{eqn:lagrangian_supp}, the transformations to the eigenbases of the one- and two-body operators (i.e., the ${\bf U}$ matrices) can be constrained by using the
definition of the matrices they diagonalize. That is, $U^\varnothing_{pk}$ contains the eigenvectors of $\mathcal{F}_{pq}$. Therefore, we can use the fact that, when tranformed to its eigenbasis, the off-diagonal matrix elements of this matrix are zero (viz. $\mathcal{F}^\varnothing_{kk'}=0$ $\forall$ $k \neq k'$). Similarly, this holds for each $t$-leaf of the two-body operator. In this case, $U^t_{pk}$ are the eigenvectors of $V^t_{pq}$ with eigenvalues $\lambda^t_{k}$. If we define a new quantity,
\begin{equation}
\lambda^t_{kk'} = \sum_{pq} U^t_{pk} V^t_{pq} U^t_{qk'},
\end{equation}
trivially, we can see that $\lambda^t_{kk'} = \lambda^t_{k}\delta_{kk'}$. However, this eigenbasis representation of $V^t_{pq}$ is useful to define the constraint on $U^t_{pk}$. We use the $\mu^\varnothing_{kk'}$ and $\mu^t_{kk'}$ undetermined multipliers to set the partial derivatives of the Lagrangian with respect to $U^\varnothing_{pk}$ and $U^t_{pk}$, respectively,
to zero. Here, one should note that the energy, the $\eta$ constraint terms and the $\mu$ constraint terms carry explicit dependence on the ${\bf U}$ matrices. Therefore, solving for the
$\mu$ undetermined multipliers is the next step after the $\eta$ multipliers have been formed.

We will begin by deriving programmable equations for $\mu^\varnothing_{kk'}$ that make the Lagrangian stationary with respect to variation in $U^\varnothing_{pk}$. To accomplish this, we will set partial derivatives of the Lagrangian with respect to $U^\varnothing_{pk}$ to zero and
transformed the resulting derivative by $U^\varnothing_{pk}$ to produce a result that is fully
in the eigenbasis of $\mathcal{F}_{pq}$. The transformation by $U^\varnothing_{pk}$ is not formally required, but will simplify the resulting equations.
\begin{align}
0 & =
\sum_{r} U^\varnothing_{rl'} \pdiff{\mathcal{L}}{U^\varnothing_{rl}}
=
\sum_{r} U^\varnothing_{rl'} \pdiff{}{U^\varnothing_{rl}}
\sum_{k}
\mathcal{F}_{k}^{\varnothing}
\omega_{k}^{\varnothing}
\nonumber \\
+
&
\sum_{r} U^\varnothing_{rl'} \pdiff{}{U^\varnothing_{rl}}
\sum_{p > k}
\eta_{pk}^{\varnothing}
\left [
\left (
\prod_{g}
\hat G^{\varnothing}
( \theta_{g}^{\varnothing} )
\right )_{pk}
-
U_{pk}^{\varnothing}
\right ] \nonumber \\
+ &
\sum_{r} U^\varnothing_{rl'} \pdiff{}{U^\varnothing_{rl}}
\sum_{k > k'}
\mu_{kk'}^{\varnothing}
\mathcal{F}_{kk'}^{\varnothing}
\end{align}
Now, we can rewrite the Lagrangian to expose the dependence on $U^\varnothing_{pk}$.
\begin{align}
0 & =
\sum_{r} U^\varnothing_{rl'} \pdiff{}{U^\varnothing_{rl}}
\sum_{k}
\sum_{pq}
U^\varnothing_{pk}
U^\varnothing_{qk}
\mathcal{F}_{pq}
\omega_{k}^{\varnothing}
\nonumber \\
+
&
\sum_{r} U^\varnothing_{rl'} \pdiff{}{U^\varnothing_{rl}}
\sum_{p > k}
\eta_{pk}^{\varnothing}
\left [
\left (
\prod_{g}
\hat G^{\varnothing}
( \theta_{g}^{\varnothing} )
\right )_{pk}
-
U_{pk}^{\varnothing}
\right ] \nonumber \\
+ &
\sum_{r} U^\varnothing_{rl'} \pdiff{}{U^\varnothing_{rl}}
\sum_{k > k'}
\sum_{pq}
\mu_{kk'}^{\varnothing}
U^\varnothing_{pk}
U^\varnothing_{qk'}
\mathcal{F}_{pq}^{\varnothing}
\end{align}
Taking the derivatives and transforming, to the eigenbasis we get:
\begin{align}
0
=
&
2\delta_{ll'}
\mathcal{F}_{l}^{\varnothing}
\omega_{l}^{\varnothing}
-
\eta_{l'l}^{\varnothing}
\nonumber \\
& +
\begin{cases}
  \mu_{ll'}^{\varnothing} \mathcal{F}_{l}^{\varnothing}
  & \text{if} \;\; l>l' \\
  \mu_{l'l}^{\varnothing} \mathcal{F}_{l}^{\varnothing}
  & \text{if} \;\; l'>l \\
\end{cases}.
\end{align}
We have transformed the molecular orbital index of the $\eta^\varnothing_{pk}$ multipliers to the eigenbasis.
\begin{equation}
\eta^\varnothing_{kk'}
=
\sum_{p}
U^\varnothing_{pk}
\eta^\varnothing_{pk'}
\end{equation}
Here, we have treated $\eta^\varnothing_{pk}$ as a square matrix containing the values of the multipliers defined by Eqn. \ref{eqn:eta0} in the lower triangle and zeros elsewhere.
Notice that we have a full matrix representing the derivatives of the Lagrangian, while we only have a lower triangular matrix of undetermined multipliers. We can define the multipliers, $\mu_{ll'}^{\varnothing}$, by equating the upper and lower triangles of the Lagrangian derivatives -- both of which must equal zero.
\begin{align}
2\delta_{ll'}
\mathcal{F}_{l}^{\varnothing}
\omega_{l}^{\varnothing}
-
\eta_{l'l}^{\varnothing}
+
&
\mu_{ll'}^{\varnothing} 
\mathcal{F}_{l}^{\varnothing}
=
\nonumber \\
&
2\delta_{ll'}
\mathcal{F}_{l}^{\varnothing}
\omega_{l}^{\varnothing}
-
\eta_{ll'}^{\varnothing}
+
\mu_{ll'}^{\varnothing} 
\mathcal{F}_{l'}^{\varnothing}
\end{align}
When we solve for the multipliers, the contributions from the eigenbasis density matrix cancel and only the $\eta_{ll'}^{\varnothing}$ contribute to the right-hand side of the system of linear equations defining
$\mu_{ll'}^{\varnothing}$.
\begin{align}
\mu_{ll'}^{\varnothing} 
= 
\frac{
\eta_{ll'}^{\varnothing}
-
\eta_{l'l}^{\varnothing}
}{\mathcal{F}_{l'}^{\varnothing} - \mathcal{F}_{l}^{\varnothing}}
\end{align}
With the $\mu_{ll'}^{\varnothing}$ multipliers defined, the
Lagrangian is stationary with respect to variation in $U_{pk}^{\varnothing}$.

Following along similar lines, we can derive an expression for $\mu^t_{kk'}$ to make the Lagrangian stationary with respect to variation in $U^t_{pk}$. Again, we start by setting derivatives of the Lagrangian with respect to $U^t_{pk}$ to zero.
\begin{align}
0 & =
\sum_{r} U^{t'}_{rm'} \pdiff{\mathcal{L}}{U^{t'}_{rm}}
=
\sum_{r} U^{t'}_{rm'} \pdiff{}{U^{t'}_{rm}}
\sum_t^T
\sum_{kl}
Z_{kl}^{t}
\omega_{kl}^{t}
\nonumber \\
+
&
\sum_{r} U^{t'}_{rm'} \pdiff{}{U^{t'}_{rm}}
\sum_t^T
\sum_{p > k}
\eta_{pk}^{t}
\left [
\left (
\prod_{g}
\hat G^{t}
( \theta_{g}^{t} )
\right )_{pk}
-
U_{pk}^{t}
\right ] \nonumber \\
+ &
\sum_{r} U^{t'}_{rm'} \pdiff{}{U^{t'}_{rm}}
\sum_t^T
\sum_{k > k'}
\mu_{kk'}^{t}
\lambda_{kk'}^{t}
\end{align}
Expanding to show explicitly the dependence on $U^t_{pk}$, we get:
\begin{align}
0 = &
\sum_{r} U^{t'}_{rm'} \pdiff{}{U^{t'}_{rm}} \Bigg(
\nonumber \\ &
\sum_t^T
\sum_{kl}
\sum_{pp'qq'}
U^t_{pk} U^t_{p'k} V^t_{pp'}
g^t
V^t_{qq'} U^t_{ql} U^t_{q'l}
\omega_{kl}^{t}
\nonumber \\ 
+
&
\sum_t^T
\sum_{p > k}
\eta_{pk}^{t}
\left [
\left (
\prod_{g}
\hat G^{t}
( \theta_{g}^{t} )
\right )_{pk}
-
U_{pk}^{t}
\right ]
\nonumber \\ 
+
&
\sum_t^T
\sum_{k > k'}
\sum_{pp'}
\mu_{kk'}^{t}
U^t_{pk} U^t_{p'k'} V^t_{pp'}
\Bigg).
\end{align}
At this point, an important observation to make is that we only need to take derivatives corresponding to $U^t_{pk}$ with $t\leq\/T$; derivatives with respect to $U^t_{pk}$ with $t>T$ are trivially zero. Now, we can take the partial derivatives and transform the result back to the eigenbasis.
\begin{align}
0 
= 
&
4 
\delta_{mm'}
\sum_k
Z^{t'}_{km}
\omega_{km}^{t'}
-
\eta_{m'm}^{t'}
\nonumber \\
& +
\begin{cases}
  \mu_{mm'}^{t'} \lambda_{m}^{t'}
  & \text{if} \;\; m>m' \\
  \mu_{m'm}^{t'} \lambda_{m}^{t'}
  & \text{if} \;\; m'>m \\
\end{cases}.
\end{align}
As we did previously, we will transform the $\eta^t_{pk'}$ multipliers into the eigenbasis.
\begin{equation}
\eta^t_{kk'}
=
\sum_{p}
U^t_{pk}
\eta^t_{pk'}
\end{equation}
Again, the $\mu_{mm'}^{t}$ multipliers are defined by a system of linear equations.
\begin{align}
4\delta_{ll'}
\sum_k
Z^{t}_{kl}
\omega_{kl}^{t}
-
&
\eta_{l'l}^{t}
+
\mu_{ll'}^{t} 
\lambda_{l}^{t}
=
\nonumber \\
&
4\delta_{ll'}
\sum_k
Z^{t}_{kl}
\omega_{kl}^{t}
-
\eta_{ll'}^{t}
+
\mu_{ll'}^{t} 
\lambda_{l'}^{t}
\end{align}
These linear equations have a simple analytic solution.
\begin{align}
\mu_{ll'}^{t} 
= 
\frac{
\eta_{ll'}^{t}
-
\eta_{l'l}^{t}
}{\lambda_{l'}^{t} - \lambda_{l}^{t}}
\end{align}
Solving for these multipliers makes the Lagrangian stationary with respect to variation in $U^t_{pk}$. In both cases ($U^\varnothing_{pk}$ and $U^t_{pk}$), the multipliers are defined separately for each $t$-leaf.

\subsection{Constraints on the Two-Electron Integral Eigendecomposition}

The $\nu^{tt'}$ multipliers depend on derivatives of the two-body contribution to the energy
as well as the $\mu^t_{pk}$ multipliers. These multipliers couple the $t$-leaves together and, in the case of truncated X-DF they account for rotations between $t$-leaves included in the definition of the Hamiltonian and those that were excluded. These multipliers make the Lagrangian stationary with respect to variation in the eigenvectors of the two-electron integrals, $V^t_{pq}$. To begin, we will set the derivatives of the Lagrangian with respect to $V^t_{pq}$ to zero.
\begin{align}
0 & =
\sum_{rr'} V^{u'}_{rr'} \pdiff{\mathcal{L}}{V^{u}_{rr'}}
=
\sum_{rr'} V^{u'}_{rr'} \pdiff{}{V^{u}_{rr'}} \Bigg(
\nonumber \\
&
\sum_{t}^{T}\sum_{kl}
Z^t_{kl}
\omega^t_{kl}
+
\sum_{t}^{T}
\sum_{k > k'}
\mu_{kk'}^{t}
\lambda_{kk'}^{t}
+
\sum_{t > t'}
\nu^{tt'}
g^{tt'}
\Bigg)
\end{align}
We must fully expand these terms to expose the explicit dependence on $V^{t}_{pq}$.
Consider  $\lambda_{kk'}^t$
\begin{align}
\lambda_{kk'}^t =
\sum_{pp'}
U^t_{pk} V^t_{pp'} U^t_{p'k'}
\end{align}
and $g^t$
\begin{align}
g^t =
\sum_{pp'qq'}
V^t_{pp'} (pp'|qq') V^t_{qq'}.
\end{align}
Some useful partial derivatives are:
\begin{align}
\sum_{rr'} V^{u'}_{rr'} \pdiff{\lambda_{kk'}^t}{V^{u}_{rr'}} =
\delta_{tu}
\sum_{rr'}
V^{u'}_{rr'}
U^t_{rk}
U^t_{r'k'}
\end{align}
and
\begin{align}
\sum_{rr'} V^{u'}_{rr'} \pdiff{g^t}{V^{u}_{rr'}} =
2 
\delta_{tu}
g^{tu'}
\end{align}
These are combined in the definition of $Z^t_{kl}$
\begin{align}
Z^t_{kl} =
\sum_{pp'qq'}
\sum_{rr'ss'}
U^t_{pk} V^t_{pp'} U^t_{p'k} 
V^t_{rr'} (rr'|ss') V^t_{ss'}
U^t_{ql} V^t_{qq'} U^t_{q'l},
\end{align}
and its partial derivatives.
\begin{align}
\sum_{rr'} V^{u'}_{rr'} & \pdiff{Z^t_{kl}}{V^{u}_{rr'}} =
2 \delta_{tu} 
\lambda_k^{t}
g^{tu'}
\lambda_l^{t}
\nonumber \\
+
&
\delta_{tu} 
\sum_{rr'}
\Big[
U^t_{rk}
V^{u'}_{rr'}
U^t_{r'k}
g^t
\lambda_l^{t}
+
\lambda_k^{t}
g^t
U^t_{rl}
V^{u'}_{rr'}
U^t_{r'l}
\Big]
\end{align}
Now, we can define a useful intermediate.
\begin{align}
R^{u'u} & = 
\sum_{rr'} V^{u'}_{rr'} \pdiff{}{V^{u}_{rr'}} \sum_t^T \Big[
\sum_{kl} Z^t_{kl} \omega_{kl}^t +
\sum_{k > k'} \mu_{kk'}^{t} \lambda_{kk'}^{t}
\Big]
\nonumber \\
& = 
2 \sum_{rr'} \sum_{kl} U^u_{rk} V^{u'}_{rr'} U^u_{r'k} g^u \lambda_l^u \omega_{kl}^u
\nonumber \\
& +
\sum_{k>k'} \sum_{rr'} \mu^u_{kk'} U^u_{rk} U^u_{r'k'} V^{u'}_{rr'}
\;\;\;\;
\forall
\;\;\;\;
u \leq T
\end{align}
\begin{align}
R^{u'u} = 0
\;\;\;\;
\forall
\;\;\;\;
u > T
\end{align}
From here, the $\nu^{tt'}$ multipliers are defined by the eigenvector response of 
$V^t_{pq}$ (as was the case for the $\mu^\varnothing_{kk'}$ and $\mu^t_{kk'}$ multipliers). As a result, the $\nu^{tt'}$ multipliers have a simple analytic form.
\begin{equation}
\nu^{tt'}
=
\frac{
R_{tt'}
-
R_{t't}
}
{
g^{t'}
-
g^t
}
\end{equation}
Once these multipliers are determined, the Lagrangian has been made stationary with respect to $V^t_{pq}$. At this point, the Lagrangian is stationary with respect to variation in any of the quantities defining the X-DF.

\subsection{Constraints on the Molecular Orbitals}

The final step before the derivative of the energy can be evaluated is to make the Lagrangian stationary with respect to variation in the underlying molecular orbitals (in this case, those determined from FON-HF). Techniques to evaluate the derivative of configuration interaction (CI) energies constructed from FON-HF orbitals are well-known \cite{Hohenstein:2015:014111} and we will not repeat the derivation here. The key to leveraging these existing approaches in the context of X-DF/VQE is to construct relaxed one- and two-body reduced density matrices within the active space. These matrices are analogous to those obtained from, for example, a complete active space CI wavefunction. These density matrices are used to define the right-hand side of the coupled-perturbed FON-HF (CP-FON-HF) equations (i.e., those that define the $z_{pq}$ multipliers). Subsequently, a combination of the density matrices and the $z_{pq}$ multipliers are used to define the $x_{pq}$ multipliers. Once these multipliers are determined, the Lagrangian has been made stationary with respect to all nonvariational parameters, and the derivative of the energy can be evaluated.

An operational definition of the one- and two-body density matrices can be obtained by differentiating the stationary Lagrangian with respect to the one- and two-body integral matrices. In what follows, it is convenient to define an identity matrix with a rank equal to the number of active orbitals,
\begin{equation}
I_{pq} = \delta_{pq}.
\end{equation}
We will define the one-body density matrix as the derivative of the Lagrangian (which has been made stationary with respect to all the X-DF non-variational parameters: $\theta_g^\varnothing$, $\theta_g^t$, $U^\varnothing_{pk}$, $U^t_{pk}$, $V^t_{pq}$) with respect to the frozen core operator, $(p|h_c|q)$. We denote this Lagrangian (which omits the $z_{pq}$ and $x_{pq}$ multipliers) as $\mathcal{L}_\text{X-DF}$.
\begin{align}
\pdiff{\mathcal{L}_\text{X-DF}}{(p|h_c|q)} & =
\nonumber \\
\gamma_{pq}
&
= I_{pq} 
+ \sum_k U^\varnothing_{pk} \omega_k^\varnothing U^\varnothing_{qk}
+ \sum_{kk'} U^\varnothing_{pk} \mu_{kk'}^\varnothing U^\varnothing_{qk'}
\end{align}
Here, the identity matrix appears as derivatives of $\mathcal{E}$. In the eigenbasis, the density matrix $\omega_k^\varnothing$ accounts for the diagonal contributions. Interestingly, we can see that the Lagrange multipliers, $\mu_{kk'}^\varnothing$, account, {\em exactly}, for the off-diagonal density matrix elements in the eigenbasis.
This illustrates how a series of nested Lagrange multipliers -- first solving for $\eta^\varnothing_{pk}$ and then for $\mu_{kk'}^\varnothing$ -- can be used to indirectly recover the full one-body density (that is, without evaluating $\gamma_{pq} = \langle \Psi | \hat{E}_{pq} | \Psi \rangle$ as would be done in a classical electronic structure program).

The two-body density matrix can be defined in a similar manner.
\begin{align}
\pdiff{\mathcal{L}_\text{X-DF}}{(pq|rs)} & =
\nonumber \\
\Gamma_{pqrs}
&
= \frac{1}{2} I_{pq} I_{rs} 
- \frac{1}{8} I_{pr} I_{qs}
- \frac{1}{8} I_{ps} I_{qr}
\nonumber \\
&
+ \bar{\gamma}_{pq} I_{rs}
- \frac{1}{4} \bar{\gamma}_{pr} I_{qs}
- \frac{1}{4} \bar{\gamma}_{ps} I_{qr}
\nonumber \\
&
+ \sum_{t}^{T}\sum_{kl} V^t_{pq} \lambda_k^t \omega^t_{kl} \lambda_l^t V^t_{rs}
+ \sum_{tu} V^t_{pq} \nu^{tu} V^u_{rs}
\end{align}
Here, we define $\gamma_{pq} = I_{pq} + \bar{\gamma}_{pq}$; this allows the two-electron contributions from the effective one-body operator, $\mathcal{F}_{pq}$, to be incorporated into the two-body density matrix. The remainder of the contributions to the two-body density matrix come from the eigenbasis density, $\omega^t_{kl}$, and the Lagrange multipliers, $\nu^{tu}$.

As defined above, the density matrices have less symmetry than may have been expected. However, in the subsequent steps involving derivatives with respect to the molecular orbital coefficients, these density matrices will be multiplied by and contracted with one- and two-body integral tensors that possess the usual 2-fold and 8-fold symmetries. As a result, we can symmetrize the density matrices to exhibit the same 2-fold and 8-fold symmetry.
\begin{align}
\tilde{\gamma}_{pq} = \frac{1}{2} 
\left( 
\gamma_{pq}
+
\gamma_{qp}
\right)
\end{align}
\begin{align}
\tilde{\Gamma}_{pqrs} = \frac{1}{8} 
\Big( 
& 
\Gamma_{pqrs}
+
\Gamma_{pqsr}
+
\Gamma_{qprs}
+
\Gamma_{qpsr}
\nonumber \\
+
&
\Gamma_{rspq}
+
\Gamma_{srpq}
+
\Gamma_{rsqp}
+
\Gamma_{srqp}
\Big)
\end{align}
These symmetrized density matrices can be used directly in existing formulation of the CP-FON-HF response equations to evaluate the $z_{pq}$ and $x_{pq}$ multipliers.

\subsection{Evaluation of the Gradient}

Once all of the Lagrange multipliers have been determined and the Lagrangian is fully stationary with respect to all non-variational parameters, the derivative of the energy can be evaluated by taking the derivative of the Lagrangian with respect to the perturbation of interest, $\xi$.
\begin{align}
\diff{E}{\xi} \equiv \diff{\mathcal{L}}{\xi} & =
\sum_{pq}
\hat{\gamma}_{pq}
(p|\hat{h}|q)^\xi
+
\sum_{pqrs}
\hat{\Gamma}_{pqrs}
(pq|rs)^\xi
\nonumber \\
& 
+ \sum_{p > q} z_{pq} f_{pq}^\xi
- \sum_{p \leq q} x_{pq} (p|q)^\xi
\end{align}
We denote the derivative integrals with superscript $\xi$, that is:
\begin{align}
(p|q)^\xi = \diff{}{\xi} (p|q) = (\diff{}{\xi}p|q) + (p|\diff{}{\xi}q)
\end{align}
Here, we have absorbed contributions from the frozen core determinant (i.e., $E_c$ and $(p|\hat{h}_c|q)$) into the $\hat{\gamma}_{pq}$ and $\hat{\Gamma}_{pqrs}$ density matrices. In many cases, it is possible to absorb the $z_{pq}$ multipliers into these density matrices as well. For generality, we contract the $z_{pq}$ multipliers against the derivative Fock matrix elements instead. This allows one to switch between molecular orbitals obtained from Hartree-Fock (which is the case where the $z_{pq}$ multipliers can be absorbed into the one- and two-body density matrices) or those from Kohn-Sham density functional theory (KS-DFT) or even those obtained from semiempirical theories. All that matters is that the coupled-perturbed response equations used to define $z_{pq}$ and $x_{pq}$ are consistent with the self-consistent field method used to determine the molecular orbitals and that the derivative Fock matrix is also consistent with that method.

\subsection{Parameter-shift rules of parameter-locked gate ensembles}

Calculating gradients of quantum operations on chip has always been a key part for variational quantum algorithms. Finite difference as a method of evaluating these is flawed due to imperfections and limited precision of measurement on these devices, so parameter shift rules \cite{PhysRevLett.118.150503,PhysRevA.98.032309,PennyLane, Mari2021,wierichs2021general} were proposed as a way of obtaining the gradient of a generative gate by evaluating the gate at specific angles and using the knowledge one has over the respective objective function. This has been extended and generalized to gates with arbitrary generators and used to not only to calculate the local derivative but extract the entire local objective function for optimisation purposes \cite{wierichs2021general}. Here we want to briefly discuss the case where one gate is composed of multiple of these gates that are parameter-locked to each other and act on distinct qubits/subspaces, as this applies to our spin-adapted  orbital rotation gate that is a composition of two Givens rotation gates acting on the alpha and beta qubits respectively.

Using the general shift-rules, one finds an 8-shift rule for the gradient of the device which seems not-optimal as it is comprised of two 2-shift gates. We can recover a 4-shift rule by unlocking the parameters for the calculation of the gradient and using the fact that the individual derivatives of the gates decouple.

To see this, quickly reminder of the derivative of the single Givens rotation:

\begin{equation}
    E(\theta) = \langle \Psi | G^\dagger(\theta) \hat{O}  G(\theta)   | \Psi \rangle
\end{equation}
\begin{equation}
    \partial_\theta E(\theta) = \langle \Psi | G^\dagger(\theta) \hat{O}  \partial_\theta G(\theta)   | \Psi \rangle + \text{h.c.}
\end{equation}
We absorbed every gate before the gate of interest into the $\langle \psi |$ and every gate after it into the observable $\hat{O}$ enforce each sub-gate to act on distinct subspaces over the commutation rules.
This can be evaluated by a two shift parameter rule. A look at the orbital rotation gate with the same premises:
\begin{equation}
    F(\theta) = \langle \Psi | G_2^\dagger(\theta) G_1^\dagger(\theta) \hat{O}  G_1(\theta) G_2(\theta)   | \Psi \rangle
\end{equation}

Evaluating the partial derivative via the product rule gives us

 \begin{align}
    \partial_\theta F(\theta) & = \langle \Psi | G_2^\dagger(\theta) G_1^\dagger(\theta) \hat{O}   G_1(\theta) \partial_{\theta'} G_2(\theta')   | \Psi \rangle \nonumber \\
    &+
    \langle \Psi | G_2^\dagger(\theta) G_1^\dagger(\theta) \hat{O}  \partial_{\theta'} G_1(\theta') G_2(\theta)   | \Psi \rangle\nonumber \\
    &+
    \text{h.c.}
\end{align}
One can now recover the two individual shift rules by rearranging and lifting the parameter-lock between the two Givens rotations to evaluate the two shift rules.

One can now see that each of these terms can now be evaluated with the parameter-shift rule of the individual gate.
This leaves us with the following 4-shift rule:
\begin{equation}
    F_\text{ul}(\theta, \theta') = \langle \Psi | G_2^\dagger(\theta') G_1^\dagger(\theta) \hat{O}  G_1(\theta) G_2(\theta')   | \Psi \rangle
\end{equation}
\begin{align}
    \partial_\theta F(\theta) & = \frac{1}{2} [ \text{F}_\text{ul}\left(\theta+\frac{\pi}{2},\theta\right)-\text{F}_\text{ul}\left(\theta-\frac{\pi}{2},\theta\right) \nonumber \\&+\text{F}_\text{ul}\left(\theta,\theta+\frac{\pi}{2}\right)-\text{F}_\text{ul}\left(\theta,\theta-\frac{\pi}{2}\right)]
\end{align}
On a sidenote, this decoupling does not allow for the construction of the local function and only recovers the local derivative at the given angle.

\section{Numerical Demonstration}

To test the correctness of the above derivation, we compute the analytic derivative of the ground state energy of butadiene. We use an active space containing 4 electrons in the 4 $\pi$ and $\pi^\ast$ orbitals. The molecular orbitals were obtained from FON-HF/6-31G** using Gaussian broadening of the orbital energy levels with a temperature parameter of 0.3 a.u. We have chosen a geometry of butadiene that is near equilibrium, but slightly distorted to break symmetry. The Cartesian coordinates of the nuclei are given below in \AA\/ngstr\"{o}ms.

\begin{verbatim}
C         25.72300       29.68500       26.90100
C         25.51000       29.88900       25.49000
C         26.07700       30.59500       27.82800
C         25.16300       28.92900       24.60400
H         25.62000       28.64500       27.18200
H         25.33300       30.87700       25.08600
H         26.26500       30.17800       28.85200
H         26.32100       31.69500       27.63100
H         24.81000       29.04000       23.56900
H         25.10400       27.88700       24.94600
\end{verbatim}

There are four cases of particular interest that we consider below. The VQE wavefunction can be converged to the exact result or it can be truncated. The X-DF can be exact (i.e., all $t$-leaves are retained) or $t$-leaves with small eigenvalues can be neglected. This leads to four cases where we will demonstrate the correctness of our analytic gradient formulation.

\subsection{Exact X-DF and Converged VQE}

The easiest case to test the correctness of our analytic gradients is where the X-DF is exact and the VQE wavefunction has been converged to the exact ground state energy. This allow direct comparison between classical complete active space configuration interaction (CASCI) analytic gradients and the X-DF/VQE analytic gradient. The numerical results are shown below in units of Hartree/Bohr.

CASCI gradient:
\begin{verbatim}
C    0.0087640    0.0338484   -0.0230318
C    0.0338931   -0.0447006    0.0585443
C   -0.0228010   -0.0204001   -0.0339556
C    0.0201028    0.0442291   -0.0012467
H   -0.0159444   -0.0116912    0.0023734
H   -0.0281456    0.0057919   -0.0161052
H    0.0108257   -0.0283045    0.0318503
H    0.0111623    0.0430530   -0.0067838
H   -0.0091125   -0.0101414   -0.0188318
H   -0.0087443   -0.0116847    0.0071869
\end{verbatim}

X-DF/VQE gradient:
\begin{verbatim}
C    0.0087640    0.0338484   -0.0230317
C    0.0338931   -0.0447006    0.0585443
C   -0.0228010   -0.0204001   -0.0339556
C    0.0201028    0.0442291   -0.0012467
H   -0.0159444   -0.0116912    0.0023734
H   -0.0281456    0.0057919   -0.0161052
H    0.0108257   -0.0283045    0.0318503
H    0.0111623    0.0430530   -0.0067838
H   -0.0091125   -0.0101414   -0.0188318
H   -0.0087443   -0.0116847    0.0071869
\end{verbatim}

In this case, we achieve agreement of $10^{-7}$ Hartree/Bohr or better in all of the nuclear degrees of freedom. This result indicates that many of the terms in our Lagrangian are correct, but it does not fully test the formulation because the X-DF has not been truncated -- all 10 of the $t$-leaves are included in the factorization.

\subsection{Truncated X-DF and Converged VQE}

It is essential that the analytic gradients be correct when the X-DF is truncated. Here, we discard all eigenvalues of the electron repulsion integral matrix that are below $10^{-1}$ a.u. This results in only 4 of 10 $t$-leaves being retained. In this case, we do not have the analogous classical implementation of the gradient available, so we compare against numerical differentiation of the X-DF/VQE energy. We use a 5-point central difference formula with a step size of $10^{-3}$ Bohr to obtain the numerical derivative.

Numerical X-DF/VQE gradient:
\begin{verbatim}
C    0.0024931    0.0253168   -0.0298848
C    0.0351976   -0.0349578    0.0654167
C   -0.0138414   -0.0122360   -0.0250024
C    0.0158476    0.0359590   -0.0099373
H   -0.0148655   -0.0104737    0.0006804
H   -0.0264798    0.0026624   -0.0136967
H    0.0100783   -0.0265528    0.0300495
H    0.0085186    0.0421931   -0.0055591
H   -0.0074004   -0.0105833   -0.0194204
H   -0.0095482   -0.0113273    0.0073544
\end{verbatim}

Analytical X-DF/VQE gradient:
\begin{verbatim}
C    0.0024931    0.0253165   -0.0298849
C    0.0351976   -0.0349580    0.0654169
C   -0.0138413   -0.0122359   -0.0250024
C    0.0158476    0.0359592   -0.0099374
H   -0.0148654   -0.0104737    0.0006803
H   -0.0264798    0.0026622   -0.0136968
H    0.0100782   -0.0265528    0.0300495
H    0.0085186    0.0421931   -0.0055592
H   -0.0074003   -0.0105833   -0.0194203
H   -0.0095482   -0.0113273    0.0073545
\end{verbatim}

Here, we achieve agreement of $10^{-6}$ Hartree/Bohr or better between analytical and numerical gradients. This indicates that our formulation of the X-DF/VQE gradient is correct when the X-DF is truncated. Had there been errors in the gradient formulation, the aggressive truncation we applied should have made them readily apparent.

\subsection{Exact X-DF and Approximate VQE}

Another case where our analytic gradient formulation should be correct is when an approximate (i.e., not converged to the exact ground state) VQE wavefunction is used. Our assumption is only that the wavefunction has been variationally optimized, not that it produces the exact energy. In this case, we again compare our analytic gradients against numerical gradients (using the same recipe as above). Note that due the the presence of many local minima in the VQE parameter space, we locate a minimum in the parameter space and use those parameters to seed the calculations at displaced geometries to avoid finding different minima at every geometry.

Numerical X-DF/VQE gradient:
\begin{verbatim}
C   -0.0018739    0.0152072   -0.0156463
C    0.0340339   -0.0220893    0.0421282
C   -0.0193957   -0.0108778   -0.0176749
C    0.0124084    0.0339842   -0.0071089
H   -0.0080573   -0.0080723   -0.0010542
H   -0.0203369   -0.0023167   -0.0109582
H    0.0086448   -0.0252825    0.0288439
H    0.0125202    0.0424496   -0.0070197
H   -0.0112387   -0.0112398   -0.0170473
H   -0.0067050   -0.0117621    0.0055379
\end{verbatim}

Analytical X-DF/VQE gradient:
\begin{verbatim}
C   -0.0018740    0.0152069   -0.0156461
C    0.0340339   -0.0220895    0.0421281
C   -0.0193956   -0.0108778   -0.0176750
C    0.0124085    0.0339845   -0.0071090
H   -0.0080573   -0.0080723   -0.0010544
H   -0.0203368   -0.0023169   -0.0109584
H    0.0086447   -0.0252825    0.0288439
H    0.0125203    0.0424496   -0.0070198
H   -0.0112387   -0.0112398   -0.0170473
H   -0.0067050   -0.0117621    0.0055379
\end{verbatim}

Once again we achieve agreement of better than $10^{-6}$ Hartree/Bohr between analytical and numerical gradients indicating the correctness of our gradient formulation.

\subsection{Truncated X-DF and Approximate VQE}

A final test is that the analytic gradients are correct in cases where both the X-DF is truncated and the VQE energy is an approximation of the exact ground state energy. For this test, we retain 4 of 10 $t$-leaves in the X-DF.

Numerical X-DF/VQE gradient:
\begin{verbatim}
C   -0.0088850    0.0045834   -0.0160591
C    0.0339181   -0.0123153    0.0518398
C   -0.0095220   -0.0020111   -0.0103695
C    0.0077376    0.0243913   -0.0198183
H   -0.0054178   -0.0065542   -0.0049439
H   -0.0184659   -0.0055151   -0.0092365
H    0.0070285   -0.0221339    0.0254742
H    0.0105200    0.0415747   -0.0049236
H   -0.0087623   -0.0114716   -0.0180051
H   -0.0081513   -0.0105478    0.0060424 
\end{verbatim}

Analytical X-DF/VQE gradient:
\begin{verbatim}
C   -0.0088851    0.0045831   -0.0160591
C    0.0339182   -0.0123156    0.0518400
C   -0.0095219   -0.0020110   -0.0103695
C    0.0077377    0.0243915   -0.0198184
H   -0.0054177   -0.0065542   -0.0049441
H   -0.0184658   -0.0055153   -0.0092367
H    0.0070284   -0.0221340    0.0254742
H    0.0105200    0.0415748   -0.0049237
H   -0.0087623   -0.0114715   -0.0180051
H   -0.0081513   -0.0105478    0.0060424
\end{verbatim}

This test shows a similar level of agreement between analytical and numerical gradients.


\begin{thebibliography}{60}%
\makeatletter
\providecommand \@ifxundefined [1]{%
 \@ifx{#1\undefined}
}%
\providecommand \@ifnum [1]{%
 \ifnum #1\expandafter \@firstoftwo
 \else \expandafter \@secondoftwo
 \fi
}%
\providecommand \@ifx [1]{%
 \ifx #1\expandafter \@firstoftwo
 \else \expandafter \@secondoftwo
 \fi
}%
\providecommand \natexlab [1]{#1}%
\providecommand \enquote  [1]{``#1''}%
\providecommand \bibnamefont  [1]{#1}%
\providecommand \bibfnamefont [1]{#1}%
\providecommand \citenamefont [1]{#1}%
\providecommand \href@noop [0]{\@secondoftwo}%
\providecommand \href [0]{\begingroup \@sanitize@url \@href}%
\providecommand \@href[1]{\@@startlink{#1}\@@href}%
\providecommand \@@href[1]{\endgroup#1\@@endlink}%
\providecommand \@sanitize@url [0]{\catcode `\\12\catcode `\$12\catcode
  `\&12\catcode `\#12\catcode `\^12\catcode `\_12\catcode `\%12\relax}%
\providecommand \@@startlink[1]{}%
\providecommand \@@endlink[0]{}%
\providecommand \url  [0]{\begingroup\@sanitize@url \@url }%
\providecommand \@url [1]{\endgroup\@href {#1}{\urlprefix }}%
\providecommand \urlprefix  [0]{URL }%
\providecommand \Eprint [0]{\href }%
\providecommand \doibase [0]{https://doi.org/}%
\providecommand \selectlanguage [0]{\@gobble}%
\providecommand \bibinfo  [0]{\@secondoftwo}%
\providecommand \bibfield  [0]{\@secondoftwo}%
\providecommand \translation [1]{[#1]}%
\providecommand \BibitemOpen [0]{}%
\providecommand \bibitemStop [0]{}%
\providecommand \bibitemNoStop [0]{.\EOS\space}%
\providecommand \EOS [0]{\spacefactor3000\relax}%
\providecommand \BibitemShut  [1]{\csname bibitem#1\endcsname}%
\let\auto@bib@innerbib\@empty
\bibitem [{\citenamefont {Bishop}\ and\ \citenamefont {Randi{\v
  c}}(1966)}]{Bishop1099}%
  \BibitemOpen
  \bibfield  {author} {\bibinfo {author} {\bibfnamefont {D.~M.}\ \bibnamefont
  {Bishop}}\ and\ \bibinfo {author} {\bibfnamefont {M.}~\bibnamefont {Randi{\v
  c}}},\ }\bibfield  {title} {\bibinfo {title} {Ab initio calculation of
  harmonic force constants},\ }\href {https://doi.org/10.1063/1.1727068}
  {\bibfield  {journal} {\bibinfo  {journal} {J. Chem. Phys.}\ }\textbf
  {\bibinfo {volume} {44}},\ \bibinfo {pages} {2480} (\bibinfo {year}
  {1966})}\BibitemShut {NoStop}%
\bibitem [{\citenamefont {Pulay}(1969)}]{Pulay1969}%
  \BibitemOpen
  \bibfield  {author} {\bibinfo {author} {\bibfnamefont {P.}~\bibnamefont
  {Pulay}},\ }\bibfield  {title} {\bibinfo {title} {Ab initio calculation of
  force constants and equilibrium geometries in polyatomic molecules},\ }\href
  {https://doi.org/10.1080/00268976900100941} {\bibfield  {journal} {\bibinfo
  {journal} {Mol. Phys.}\ }\textbf {\bibinfo {volume} {17}},\ \bibinfo {pages}
  {197} (\bibinfo {year} {1969})}\BibitemShut {NoStop}%
\bibitem [{\citenamefont {Yamaguchi}\ \emph {et~al.}(1994)\citenamefont
  {Yamaguchi}, \citenamefont {Osamura}, \citenamefont {Goddard},\ and\
  \citenamefont {Schaefer}}]{Yamaguchi:1994}%
  \BibitemOpen
  \bibfield  {author} {\bibinfo {author} {\bibfnamefont {Y.}~\bibnamefont
  {Yamaguchi}}, \bibinfo {author} {\bibfnamefont {Y.}~\bibnamefont {Osamura}},
  \bibinfo {author} {\bibfnamefont {J.~D.}\ \bibnamefont {Goddard}},\ and\
  \bibinfo {author} {\bibfnamefont {H.~F.}\ \bibnamefont {Schaefer}},\ }\href
  {https://doi.org/10.1021/ed072pA42.6} {\emph {\bibinfo {title} {A New
  Dimension to Quantum Chemistry: {Analytic} Derivative Methods in Ab Initio
  Molecular Electronic Structure Theory}}}\ (\bibinfo  {publisher} {Oxford
  University Press},\ \bibinfo {address} {Oxford},\ \bibinfo {year}
  {1994})\BibitemShut {NoStop}%
\bibitem [{\citenamefont {Pulay}(2013)}]{Pulay2014}%
  \BibitemOpen
  \bibfield  {author} {\bibinfo {author} {\bibfnamefont {P.}~\bibnamefont
  {Pulay}},\ }\bibfield  {title} {\bibinfo {title} {Analytical derivatives,
  forces, force constants, molecular geometries, and related response
  properties in electronic structure theory},\ }\href
  {https://doi.org/10.1002/wcms.1171} {\bibfield  {journal} {\bibinfo
  {journal} {WIREs Comput. Mol. Sci.}\ }\textbf {\bibinfo {volume} {4}},\
  \bibinfo {pages} {169} (\bibinfo {year} {2013})}\BibitemShut {NoStop}%
\bibitem [{\citenamefont {Jensen}(2017)}]{Jensen:2017}%
  \BibitemOpen
  \bibfield  {author} {\bibinfo {author} {\bibfnamefont {F.}~\bibnamefont
  {Jensen}},\ }\href@noop {} {\emph {\bibinfo {title} {Introduction to
  computational chemistry}}}\ (\bibinfo  {publisher} {John Wiley \& sons},\
  \bibinfo {year} {2017})\BibitemShut {NoStop}%
\bibitem [{\citenamefont {Handy}\ and\ \citenamefont
  {Schaefer}(1984)}]{handy1984evaluation}%
  \BibitemOpen
  \bibfield  {author} {\bibinfo {author} {\bibfnamefont {N.~C.}\ \bibnamefont
  {Handy}}\ and\ \bibinfo {author} {\bibfnamefont {H.~F.}\ \bibnamefont
  {Schaefer}},\ }\bibfield  {title} {\bibinfo {title} {On the evaluation of
  analytic energy derivatives for correlated wave functions},\ }\href
  {https://doi.org/10.1063/1.447489} {\bibfield  {journal} {\bibinfo  {journal}
  {J. Chem. Phys.}\ }\textbf {\bibinfo {volume} {81}},\ \bibinfo {pages} {5031}
  (\bibinfo {year} {1984})}\BibitemShut {NoStop}%
\bibitem [{\citenamefont {Peruzzo}\ \emph {et~al.}(2014)\citenamefont
  {Peruzzo}, \citenamefont {McClean}, \citenamefont {Shadbolt}, \citenamefont
  {Yung}, \citenamefont {Zhou}, \citenamefont {Love}, \citenamefont
  {Aspuru-Guzik},\ and\ \citenamefont
  {O{\textquoteright}Brien}}]{Peruzzo:2014:1}%
  \BibitemOpen
  \bibfield  {author} {\bibinfo {author} {\bibfnamefont {A.}~\bibnamefont
  {Peruzzo}}, \bibinfo {author} {\bibfnamefont {J.}~\bibnamefont {McClean}},
  \bibinfo {author} {\bibfnamefont {P.}~\bibnamefont {Shadbolt}}, \bibinfo
  {author} {\bibfnamefont {M.-H.}\ \bibnamefont {Yung}}, \bibinfo {author}
  {\bibfnamefont {X.-Q.}\ \bibnamefont {Zhou}}, \bibinfo {author}
  {\bibfnamefont {P.~J.}\ \bibnamefont {Love}}, \bibinfo {author}
  {\bibfnamefont {A.}~\bibnamefont {Aspuru-Guzik}},\ and\ \bibinfo {author}
  {\bibfnamefont {J.~L.}\ \bibnamefont {O{\textquoteright}Brien}},\ }\bibfield
  {title} {\bibinfo {title} {A variational eigenvalue solver on a photonic
  quantum processor},\ }\href {https://doi.org/10.1038/ncomms5213} {\bibfield
  {journal} {\bibinfo  {journal} {Nat. Commun.}\ }\textbf {\bibinfo {volume}
  {5}},\ \bibinfo {pages} {1} (\bibinfo {year} {2014})}\BibitemShut {NoStop}%
\bibitem [{\citenamefont {McClean}\ \emph {et~al.}(2016)\citenamefont
  {McClean}, \citenamefont {Romero}, \citenamefont {Babbush},\ and\
  \citenamefont {Aspuru-Guzik}}]{McClean:2016:023023}%
  \BibitemOpen
  \bibfield  {author} {\bibinfo {author} {\bibfnamefont {J.~R.}\ \bibnamefont
  {McClean}}, \bibinfo {author} {\bibfnamefont {J.}~\bibnamefont {Romero}},
  \bibinfo {author} {\bibfnamefont {R.}~\bibnamefont {Babbush}},\ and\ \bibinfo
  {author} {\bibfnamefont {A.}~\bibnamefont {Aspuru-Guzik}},\ }\bibfield
  {title} {\bibinfo {title} {The theory of variational hybrid quantum-classical
  algorithms},\ }\href {https://doi.org/10.1088/1367-2630/18/2/023023}
  {\bibfield  {journal} {\bibinfo  {journal} {New J. Phys.}\ }\textbf {\bibinfo
  {volume} {18}},\ \bibinfo {pages} {023023} (\bibinfo {year}
  {2016})}\BibitemShut {NoStop}%
\bibitem [{\citenamefont {Romero}\ \emph {et~al.}(2018)\citenamefont {Romero},
  \citenamefont {Babbush}, \citenamefont {McClean}, \citenamefont {Hempel},
  \citenamefont {Love},\ and\ \citenamefont
  {Aspuru-Guzik}}]{Romero:2018:014008}%
  \BibitemOpen
  \bibfield  {author} {\bibinfo {author} {\bibfnamefont {J.}~\bibnamefont
  {Romero}}, \bibinfo {author} {\bibfnamefont {R.}~\bibnamefont {Babbush}},
  \bibinfo {author} {\bibfnamefont {J.~R.}\ \bibnamefont {McClean}}, \bibinfo
  {author} {\bibfnamefont {C.}~\bibnamefont {Hempel}}, \bibinfo {author}
  {\bibfnamefont {P.~J.}\ \bibnamefont {Love}},\ and\ \bibinfo {author}
  {\bibfnamefont {A.}~\bibnamefont {Aspuru-Guzik}},\ }\bibfield  {title}
  {\bibinfo {title} {Strategies for quantum computing molecular energies using
  the unitary coupled cluster ansatz},\ }\href
  {https://doi.org/10.1088/2058-9565/aad3e4} {\bibfield  {journal} {\bibinfo
  {journal} {Quantum Sci. Technol.}\ }\textbf {\bibinfo {volume} {4}},\
  \bibinfo {pages} {014008} (\bibinfo {year} {2018})}\BibitemShut {NoStop}%
\bibitem [{\citenamefont {Rice}\ \emph {et~al.}(1986)\citenamefont {Rice},
  \citenamefont {Amos}, \citenamefont {Handy}, \citenamefont {Lee},\ and\
  \citenamefont {Schaefer}}]{rice1986analytic}%
  \BibitemOpen
  \bibfield  {author} {\bibinfo {author} {\bibfnamefont {J.~E.}\ \bibnamefont
  {Rice}}, \bibinfo {author} {\bibfnamefont {R.~D.}\ \bibnamefont {Amos}},
  \bibinfo {author} {\bibfnamefont {N.~C.}\ \bibnamefont {Handy}}, \bibinfo
  {author} {\bibfnamefont {T.~J.}\ \bibnamefont {Lee}},\ and\ \bibinfo {author}
  {\bibfnamefont {H.~F.}\ \bibnamefont {Schaefer}},\ }\bibfield  {title}
  {\bibinfo {title} {The analytic configuration interaction gradient method:
  Application to the cyclic and open isomers of the {S}{$_3$} molecule},\
  }\href {https://doi.org/10.1063/1.451253} {\bibfield  {journal} {\bibinfo
  {journal} {J. Chem. Phys.}\ }\textbf {\bibinfo {volume} {85}},\ \bibinfo
  {pages} {963} (\bibinfo {year} {1986})}\BibitemShut {NoStop}%
\bibitem [{\citenamefont {Helgaker}(1982)}]{Helgaker:1982:939}%
  \BibitemOpen
  \bibfield  {author} {\bibinfo {author} {\bibfnamefont {T.~U.}\ \bibnamefont
  {Helgaker}},\ }\bibfield  {title} {\bibinfo {title} {Simple derivation of the
  potential energy gradient for an arbitrary electronic wave function},\ }\href
  {https://doi.org/10.1002/qua.560210520} {\bibfield  {journal} {\bibinfo
  {journal} {Int. J. Quantum Chem.}\ }\textbf {\bibinfo {volume} {21}},\
  \bibinfo {pages} {939} (\bibinfo {year} {1982})}\BibitemShut {NoStop}%
\bibitem [{\citenamefont {Kivlichan}\ \emph {et~al.}(2018)\citenamefont
  {Kivlichan}, \citenamefont {McClean}, \citenamefont {Wiebe}, \citenamefont
  {Gidney}, \citenamefont {Aspuru-Guzik}, \citenamefont {Chan},\ and\
  \citenamefont {Babbush}}]{PhysRevLett.120.110501}%
  \BibitemOpen
  \bibfield  {author} {\bibinfo {author} {\bibfnamefont {I.~D.}\ \bibnamefont
  {Kivlichan}}, \bibinfo {author} {\bibfnamefont {J.}~\bibnamefont {McClean}},
  \bibinfo {author} {\bibfnamefont {N.}~\bibnamefont {Wiebe}}, \bibinfo
  {author} {\bibfnamefont {C.}~\bibnamefont {Gidney}}, \bibinfo {author}
  {\bibfnamefont {A.}~\bibnamefont {Aspuru-Guzik}}, \bibinfo {author}
  {\bibfnamefont {G.~K.-L.}\ \bibnamefont {Chan}},\ and\ \bibinfo {author}
  {\bibfnamefont {R.}~\bibnamefont {Babbush}},\ }\bibfield  {title} {\bibinfo
  {title} {Quantum simulation of electronic structure with linear depth and
  connectivity},\ }\href {https://doi.org/10.1103/physrevlett.120.110501}
  {\bibfield  {journal} {\bibinfo  {journal} {Phys. Rev. Lett.}\ }\textbf
  {\bibinfo {volume} {120}},\ \bibinfo {pages} {110501} (\bibinfo {year}
  {2018})}\BibitemShut {NoStop}%
\bibitem [{\citenamefont {Berry}\ \emph {et~al.}(2019)\citenamefont {Berry},
  \citenamefont {Gidney}, \citenamefont {Motta}, \citenamefont {McClean},\ and\
  \citenamefont {Babbush}}]{Berry:2019:208}%
  \BibitemOpen
  \bibfield  {author} {\bibinfo {author} {\bibfnamefont {D.~W.}\ \bibnamefont
  {Berry}}, \bibinfo {author} {\bibfnamefont {C.}~\bibnamefont {Gidney}},
  \bibinfo {author} {\bibfnamefont {M.}~\bibnamefont {Motta}}, \bibinfo
  {author} {\bibfnamefont {J.~R.}\ \bibnamefont {McClean}},\ and\ \bibinfo
  {author} {\bibfnamefont {R.}~\bibnamefont {Babbush}},\ }\bibfield  {title}
  {\bibinfo {title} {Qubitization of arbitrary basis quantum chemistry
  leveraging sparsity and low rank factorization},\ }\href
  {https://doi.org/10.22331/q-2019-12-02-208} {\bibfield  {journal} {\bibinfo
  {journal} {Quantum}\ }\textbf {\bibinfo {volume} {3}},\ \bibinfo {pages}
  {208} (\bibinfo {year} {2019})}\BibitemShut {NoStop}%
\bibitem [{\citenamefont {Huggins}\ \emph {et~al.}(2021)\citenamefont
  {Huggins}, \citenamefont {McClean}, \citenamefont {Rubin}, \citenamefont
  {Jiang}, \citenamefont {Wiebe}, \citenamefont {Whaley},\ and\ \citenamefont
  {Babbush}}]{Huggins2021}%
  \BibitemOpen
  \bibfield  {author} {\bibinfo {author} {\bibfnamefont {W.~J.}\ \bibnamefont
  {Huggins}}, \bibinfo {author} {\bibfnamefont {J.~R.}\ \bibnamefont
  {McClean}}, \bibinfo {author} {\bibfnamefont {N.~C.}\ \bibnamefont {Rubin}},
  \bibinfo {author} {\bibfnamefont {Z.}~\bibnamefont {Jiang}}, \bibinfo
  {author} {\bibfnamefont {N.}~\bibnamefont {Wiebe}}, \bibinfo {author}
  {\bibfnamefont {K.~B.}\ \bibnamefont {Whaley}},\ and\ \bibinfo {author}
  {\bibfnamefont {R.}~\bibnamefont {Babbush}},\ }\bibfield  {title} {\bibinfo
  {title} {Efficient and noise resilient measurements for quantum chemistry on
  near-term quantum computers},\ }\bibfield  {journal} {\bibinfo  {journal}
  {npj Quantum Inf.}\ }\textbf {\bibinfo {volume} {7}},\ \href
  {https://doi.org/10.1038/s41534-020-00341-7} {10.1038/s41534-020-00341-7}
  (\bibinfo {year} {2021})\BibitemShut {NoStop}%
\bibitem [{\citenamefont {Motta}\ \emph {et~al.}(2021)\citenamefont {Motta},
  \citenamefont {Ye}, \citenamefont {McClean}, \citenamefont {Li},
  \citenamefont {Minnich}, \citenamefont {Babbush},\ and\ \citenamefont
  {Chan}}]{Motta_2021}%
  \BibitemOpen
  \bibfield  {author} {\bibinfo {author} {\bibfnamefont {M.}~\bibnamefont
  {Motta}}, \bibinfo {author} {\bibfnamefont {E.}~\bibnamefont {Ye}}, \bibinfo
  {author} {\bibfnamefont {J.~R.}\ \bibnamefont {McClean}}, \bibinfo {author}
  {\bibfnamefont {Z.}~\bibnamefont {Li}}, \bibinfo {author} {\bibfnamefont
  {A.~J.}\ \bibnamefont {Minnich}}, \bibinfo {author} {\bibfnamefont
  {R.}~\bibnamefont {Babbush}},\ and\ \bibinfo {author} {\bibfnamefont
  {G.~K.-L.}\ \bibnamefont {Chan}},\ }\bibfield  {title} {\bibinfo {title} {Low
  rank representations for quantum simulation of electronic structure},\
  }\bibfield  {journal} {\bibinfo  {journal} {npj Quantum Inf.}\ }\textbf
  {\bibinfo {volume} {7}},\ \href {https://doi.org/10.1038/s41534-021-00416-z}
  {10.1038/s41534-021-00416-z} (\bibinfo {year} {2021})\BibitemShut {NoStop}%
\bibitem [{\citenamefont {Cohn}\ \emph {et~al.}(2021)\citenamefont {Cohn},
  \citenamefont {Motta},\ and\ \citenamefont {Parrish}}]{Cohn2021filter}%
  \BibitemOpen
  \bibfield  {author} {\bibinfo {author} {\bibfnamefont {J.}~\bibnamefont
  {Cohn}}, \bibinfo {author} {\bibfnamefont {M.}~\bibnamefont {Motta}},\ and\
  \bibinfo {author} {\bibfnamefont {R.~M.}\ \bibnamefont {Parrish}},\
  }\bibfield  {title} {\bibinfo {title} {Quantum filter diagonalization with
  compressed double-factorized hamiltonians},\ }\bibfield  {journal} {\bibinfo
  {journal} {PRX Quantum}\ }\textbf {\bibinfo {volume} {2}},\ \href
  {https://doi.org/10.1103/prxquantum.2.040352} {10.1103/prxquantum.2.040352}
  (\bibinfo {year} {2021})\BibitemShut {NoStop}%
\bibitem [{\citenamefont {Vahtras}\ \emph {et~al.}(1993)\citenamefont
  {Vahtras}, \citenamefont {Alml\"of},\ and\ \citenamefont
  {Feyereisen}}]{Vahtras:1993:514}%
  \BibitemOpen
  \bibfield  {author} {\bibinfo {author} {\bibfnamefont {O.}~\bibnamefont
  {Vahtras}}, \bibinfo {author} {\bibfnamefont {J.}~\bibnamefont {Alml\"of}},\
  and\ \bibinfo {author} {\bibfnamefont {M.}~\bibnamefont {Feyereisen}},\
  }\bibfield  {title} {\bibinfo {title} {Integral approximations for
  {LCAO\}-\{SCF} calculations},\ }\href
  {https://doi.org/10.1016/0009-2614(93)89151-7} {\bibfield  {journal}
  {\bibinfo  {journal} {Chem. Phys. Lett.}\ }\textbf {\bibinfo {volume}
  {213}},\ \bibinfo {pages} {514} (\bibinfo {year} {1993})}\BibitemShut
  {NoStop}%
\bibitem [{\citenamefont {Feyereisen}\ \emph {et~al.}(1993)\citenamefont
  {Feyereisen}, \citenamefont {Fitzgerald},\ and\ \citenamefont
  {Komornicki}}]{Feyereisen:1993:359}%
  \BibitemOpen
  \bibfield  {author} {\bibinfo {author} {\bibfnamefont {M.}~\bibnamefont
  {Feyereisen}}, \bibinfo {author} {\bibfnamefont {G.}~\bibnamefont
  {Fitzgerald}},\ and\ \bibinfo {author} {\bibfnamefont {A.}~\bibnamefont
  {Komornicki}},\ }\bibfield  {title} {\bibinfo {title} {Use of approximate
  integrals in ab initio theory. an application in {MP2} energy calculations},\
  }\href {https://doi.org/10.1016/0009-2614(93)87156-w} {\bibfield  {journal}
  {\bibinfo  {journal} {Chem. Phys. Lett.}\ }\textbf {\bibinfo {volume}
  {208}},\ \bibinfo {pages} {359} (\bibinfo {year} {1993})}\BibitemShut
  {NoStop}%
\bibitem [{\citenamefont {Beebe}\ and\ \citenamefont
  {Linderberg}(1977)}]{Beebe:1977:683}%
  \BibitemOpen
  \bibfield  {author} {\bibinfo {author} {\bibfnamefont {N.~H.~F.}\
  \bibnamefont {Beebe}}\ and\ \bibinfo {author} {\bibfnamefont
  {J.}~\bibnamefont {Linderberg}},\ }\bibfield  {title} {\bibinfo {title}
  {Simplifications in the generation and transformation of two-electron
  integrals in molecular calculations},\ }\href
  {https://doi.org/10.1002/qua.560120408} {\bibfield  {journal} {\bibinfo
  {journal} {Int. J. Quantum Chem.}\ }\textbf {\bibinfo {volume} {12}},\
  \bibinfo {pages} {683} (\bibinfo {year} {1977})}\BibitemShut {NoStop}%
\bibitem [{\citenamefont {R{\o}eggen}\ and\ \citenamefont
  {Wisl{\o}ff-Nilssen}(1986)}]{Roeggen:1986:154}%
  \BibitemOpen
  \bibfield  {author} {\bibinfo {author} {\bibfnamefont {I.}~\bibnamefont
  {R{\o}eggen}}\ and\ \bibinfo {author} {\bibfnamefont {E.}~\bibnamefont
  {Wisl{\o}ff-Nilssen}},\ }\bibfield  {title} {\bibinfo {title} {On the
  \{Beebe\}-linderberg two-electron integral approximation},\ }\href
  {https://doi.org/10.1016/0009-2614(86)80099-9} {\bibfield  {journal}
  {\bibinfo  {journal} {Chem. Phys. Lett.}\ }\textbf {\bibinfo {volume}
  {132}},\ \bibinfo {pages} {154} (\bibinfo {year} {1986})}\BibitemShut
  {NoStop}%
\bibitem [{\citenamefont {Koch}\ \emph {et~al.}(2003)\citenamefont {Koch},
  \citenamefont {S\'anchez~de Mer\'as},\ and\ \citenamefont
  {Pedersen}}]{Koch:2003:9481}%
  \BibitemOpen
  \bibfield  {author} {\bibinfo {author} {\bibfnamefont {H.}~\bibnamefont
  {Koch}}, \bibinfo {author} {\bibfnamefont {A.}~\bibnamefont {S\'anchez~de
  Mer\'as}},\ and\ \bibinfo {author} {\bibfnamefont {T.~B.}\ \bibnamefont
  {Pedersen}},\ }\bibfield  {title} {\bibinfo {title} {Reduced scaling in
  electronic structure calculations using cholesky decompositions},\ }\href
  {https://doi.org/10.1063/1.1578621} {\bibfield  {journal} {\bibinfo
  {journal} {J. Chem. Phys.}\ }\textbf {\bibinfo {volume} {118}},\ \bibinfo
  {pages} {9481} (\bibinfo {year} {2003})}\BibitemShut {NoStop}%
\bibitem [{\citenamefont {Friesner}(1986)}]{Friesner:1986:1462}%
  \BibitemOpen
  \bibfield  {author} {\bibinfo {author} {\bibfnamefont {R.~A.}\ \bibnamefont
  {Friesner}},\ }\bibfield  {title} {\bibinfo {title} {Solution of the
  {Hartree{\textendash}Fock} equations by a pseudospectral method:
  {Application} to diatomic molecules},\ }\href
  {https://doi.org/10.1063/1.451237} {\bibfield  {journal} {\bibinfo  {journal}
  {J. Chem. Phys.}\ }\textbf {\bibinfo {volume} {85}},\ \bibinfo {pages} {1462}
  (\bibinfo {year} {1986})}\BibitemShut {NoStop}%
\bibitem [{\citenamefont {Friesner}(1987)}]{Friesner:1987:3522}%
  \BibitemOpen
  \bibfield  {author} {\bibinfo {author} {\bibfnamefont {R.~A.}\ \bibnamefont
  {Friesner}},\ }\bibfield  {title} {\bibinfo {title} {Solution of the
  {Hartree{\textendash}Fock} equations for polyatomic molecules by a
  pseudospectral method},\ }\href {https://doi.org/10.1063/1.451955} {\bibfield
   {journal} {\bibinfo  {journal} {J. Chem. Phys.}\ }\textbf {\bibinfo {volume}
  {86}},\ \bibinfo {pages} {3522} (\bibinfo {year} {1987})}\BibitemShut
  {NoStop}%
\bibitem [{\citenamefont {Friesner}(1991)}]{Friesner:1991:341}%
  \BibitemOpen
  \bibfield  {author} {\bibinfo {author} {\bibfnamefont {R.~A.}\ \bibnamefont
  {Friesner}},\ }\bibfield  {title} {\bibinfo {title} {New methods for
  electronic structure calculations on large molecules},\ }\href
  {https://doi.org/10.1146/annurev.pc.42.100191.002013} {\bibfield  {journal}
  {\bibinfo  {journal} {Annu. Rev. Phys. Chem.}\ }\textbf {\bibinfo {volume}
  {42}},\ \bibinfo {pages} {341} (\bibinfo {year} {1991})}\BibitemShut
  {NoStop}%
\bibitem [{\citenamefont {Izs\'ak}\ and\ \citenamefont
  {Neese}(2011)}]{Izsak:2011:144105}%
  \BibitemOpen
  \bibfield  {author} {\bibinfo {author} {\bibfnamefont {R.}~\bibnamefont
  {Izs\'ak}}\ and\ \bibinfo {author} {\bibfnamefont {F.}~\bibnamefont
  {Neese}},\ }\bibfield  {title} {\bibinfo {title} {An overlap fitted chain of
  spheres exchange method},\ }\href {https://doi.org/10.1063/1.3646921}
  {\bibfield  {journal} {\bibinfo  {journal} {J. Chem. Phys.}\ }\textbf
  {\bibinfo {volume} {135}},\ \bibinfo {pages} {144105} (\bibinfo {year}
  {2011})}\BibitemShut {NoStop}%
\bibitem [{\citenamefont {Hohenstein}\ \emph {et~al.}(2012)\citenamefont
  {Hohenstein}, \citenamefont {Parrish},\ and\ \citenamefont
  {Mart{\'\i}nez}}]{Hohenstein:2012:044103}%
  \BibitemOpen
  \bibfield  {author} {\bibinfo {author} {\bibfnamefont {E.~G.}\ \bibnamefont
  {Hohenstein}}, \bibinfo {author} {\bibfnamefont {R.~M.}\ \bibnamefont
  {Parrish}},\ and\ \bibinfo {author} {\bibfnamefont {T.~J.}\ \bibnamefont
  {Mart{\'\i}nez}},\ }\bibfield  {title} {\bibinfo {title} {Tensor
  hypercontraction density fitting. i. {Quartic} scaling second- and
  third-order {M}{\o}ller-{Plesset} perturbation theory},\ }\href
  {https://doi.org/10.1063/1.4732310} {\bibfield  {journal} {\bibinfo
  {journal} {J. Chem. Phys.}\ }\textbf {\bibinfo {volume} {137}},\ \bibinfo
  {pages} {044103} (\bibinfo {year} {2012})}\BibitemShut {NoStop}%
\bibitem [{\citenamefont {Parrish}\ \emph {et~al.}(2012)\citenamefont
  {Parrish}, \citenamefont {Hohenstein}, \citenamefont {Mart{\'\i}nez},\ and\
  \citenamefont {Sherrill}}]{Parrish:2012:224106}%
  \BibitemOpen
  \bibfield  {author} {\bibinfo {author} {\bibfnamefont {R.~M.}\ \bibnamefont
  {Parrish}}, \bibinfo {author} {\bibfnamefont {E.~G.}\ \bibnamefont
  {Hohenstein}}, \bibinfo {author} {\bibfnamefont {T.~J.}\ \bibnamefont
  {Mart{\'\i}nez}},\ and\ \bibinfo {author} {\bibfnamefont {C.~D.}\
  \bibnamefont {Sherrill}},\ }\bibfield  {title} {\bibinfo {title} {Tensor
  hypercontraction. {II.} {Least}-squares renormalization},\ }\href
  {https://doi.org/10.1063/1.4768233} {\bibfield  {journal} {\bibinfo
  {journal} {J. Chem. Phys.}\ }\textbf {\bibinfo {volume} {137}},\ \bibinfo
  {pages} {224106} (\bibinfo {year} {2012})}\BibitemShut {NoStop}%
\bibitem [{\citenamefont {Parrish}\ \emph {et~al.}(2013)\citenamefont
  {Parrish}, \citenamefont {Hohenstein}, \citenamefont {Schunck}, \citenamefont
  {Sherrill},\ and\ \citenamefont {Mart{\'\i}nez}}]{Parrish:2013:132505}%
  \BibitemOpen
  \bibfield  {author} {\bibinfo {author} {\bibfnamefont {R.~M.}\ \bibnamefont
  {Parrish}}, \bibinfo {author} {\bibfnamefont {E.~G.}\ \bibnamefont
  {Hohenstein}}, \bibinfo {author} {\bibfnamefont {N.~F.}\ \bibnamefont
  {Schunck}}, \bibinfo {author} {\bibfnamefont {C.~D.}\ \bibnamefont
  {Sherrill}},\ and\ \bibinfo {author} {\bibfnamefont {T.~J.}\ \bibnamefont
  {Mart{\'\i}nez}},\ }\bibfield  {title} {\bibinfo {title} {Exact tensor
  hypercontraction: A universal technique for the resolution of matrix elements
  of local finite-range {N}-body potentials in many-body quantum problems},\
  }\href {https://doi.org/10.1103/physrevlett.111.132505} {\bibfield  {journal}
  {\bibinfo  {journal} {Phys. Rev. Lett.}\ }\textbf {\bibinfo {volume} {111}},\
  \bibinfo {pages} {132505} (\bibinfo {year} {2013})}\BibitemShut {NoStop}%
\bibitem [{\citenamefont {Distasio}\ \emph {et~al.}(2007)\citenamefont
  {Distasio}, \citenamefont {Steele}, \citenamefont {Rhee}, \citenamefont
  {Shao},\ and\ \citenamefont {Head-Gordon}}]{Distasio:2007:839}%
  \BibitemOpen
  \bibfield  {author} {\bibinfo {author} {\bibfnamefont {R.~A.}\ \bibnamefont
  {Distasio}}, \bibinfo {author} {\bibfnamefont {R.~P.}\ \bibnamefont
  {Steele}}, \bibinfo {author} {\bibfnamefont {Y.~M.}\ \bibnamefont {Rhee}},
  \bibinfo {author} {\bibfnamefont {Y.}~\bibnamefont {Shao}},\ and\ \bibinfo
  {author} {\bibfnamefont {M.}~\bibnamefont {Head-Gordon}},\ }\bibfield
  {title} {\bibinfo {title} {An improved algorithm for analytical gradient
  evaluation in resolution-of-the-identity second-order m{\o}ller-plesset
  perturbation theory: {Application} to alanine tetrapeptide conformational
  analysis},\ }\href {https://doi.org/10.1002/jcc.20604} {\bibfield  {journal}
  {\bibinfo  {journal} {J. Comput. Chem.}\ }\textbf {\bibinfo {volume} {28}},\
  \bibinfo {pages} {839} (\bibinfo {year} {2007})}\BibitemShut {NoStop}%
\bibitem [{\citenamefont {Bostr\"om}\ \emph {et~al.}(2012)\citenamefont
  {Bostr\"om}, \citenamefont {Aquilante}, \citenamefont {Pedersen},\ and\
  \citenamefont {Lindh}}]{Bostrom:2013:204}%
  \BibitemOpen
  \bibfield  {author} {\bibinfo {author} {\bibfnamefont {J.}~\bibnamefont
  {Bostr\"om}}, \bibinfo {author} {\bibfnamefont {F.}~\bibnamefont
  {Aquilante}}, \bibinfo {author} {\bibfnamefont {T.~B.}\ \bibnamefont
  {Pedersen}},\ and\ \bibinfo {author} {\bibfnamefont {R.}~\bibnamefont
  {Lindh}},\ }\bibfield  {title} {\bibinfo {title} {Analytical gradients of
  {Hartree{\textendash}Fock} exchange with density fitting approximations},\
  }\href {https://doi.org/10.1021/ct200836x} {\bibfield  {journal} {\bibinfo
  {journal} {J. Chem. Theory Comput.}\ }\textbf {\bibinfo {volume} {9}},\
  \bibinfo {pages} {204} (\bibinfo {year} {2012})}\BibitemShut {NoStop}%
\bibitem [{\citenamefont {Delcey}\ \emph {et~al.}(2015)\citenamefont {Delcey},
  \citenamefont {Pedersen}, \citenamefont {Aquilante},\ and\ \citenamefont
  {Lindh}}]{Delcey:2015:044110}%
  \BibitemOpen
  \bibfield  {author} {\bibinfo {author} {\bibfnamefont {M.~G.}\ \bibnamefont
  {Delcey}}, \bibinfo {author} {\bibfnamefont {T.~B.}\ \bibnamefont
  {Pedersen}}, \bibinfo {author} {\bibfnamefont {F.}~\bibnamefont
  {Aquilante}},\ and\ \bibinfo {author} {\bibfnamefont {R.}~\bibnamefont
  {Lindh}},\ }\bibfield  {title} {\bibinfo {title} {Analytical gradients of the
  state-average complete active space self-consistent field method with density
  fitting},\ }\href {https://doi.org/10.1063/1.4927228} {\bibfield  {journal}
  {\bibinfo  {journal} {J. Chem. Phys.}\ }\textbf {\bibinfo {volume} {143}},\
  \bibinfo {pages} {044110} (\bibinfo {year} {2015})}\BibitemShut {NoStop}%
\bibitem [{\citenamefont {Aquilante}\ \emph {et~al.}(2008)\citenamefont
  {Aquilante}, \citenamefont {Lindh},\ and\ \citenamefont
  {Pedersen}}]{Aquilante:2008:034106}%
  \BibitemOpen
  \bibfield  {author} {\bibinfo {author} {\bibfnamefont {F.}~\bibnamefont
  {Aquilante}}, \bibinfo {author} {\bibfnamefont {R.}~\bibnamefont {Lindh}},\
  and\ \bibinfo {author} {\bibfnamefont {T.~B.}\ \bibnamefont {Pedersen}},\
  }\bibfield  {title} {\bibinfo {title} {Analytic derivatives for the cholesky
  representation of the two-electron integrals},\ }\href
  {https://doi.org/10.1063/1.2955755} {\bibfield  {journal} {\bibinfo
  {journal} {J. Chem. Phys.}\ }\textbf {\bibinfo {volume} {129}},\ \bibinfo
  {pages} {034106} (\bibinfo {year} {2008})}\BibitemShut {NoStop}%
\bibitem [{\citenamefont {Aquilante}\ \emph {et~al.}(2009)\citenamefont
  {Aquilante}, \citenamefont {Gagliardi}, \citenamefont {Pedersen},\ and\
  \citenamefont {Lindh}}]{Aquilante:2009:154107}%
  \BibitemOpen
  \bibfield  {author} {\bibinfo {author} {\bibfnamefont {F.}~\bibnamefont
  {Aquilante}}, \bibinfo {author} {\bibfnamefont {L.}~\bibnamefont
  {Gagliardi}}, \bibinfo {author} {\bibfnamefont {T.~B.}\ \bibnamefont
  {Pedersen}},\ and\ \bibinfo {author} {\bibfnamefont {R.}~\bibnamefont
  {Lindh}},\ }\bibfield  {title} {\bibinfo {title} {Atomic cholesky
  decompositions: {A} route to unbiased auxiliary basis sets for density
  fitting approximation with tunable accuracy and efficiency},\ }\href
  {https://doi.org/10.1063/1.3116784} {\bibfield  {journal} {\bibinfo
  {journal} {J. Chem. Phys.}\ }\textbf {\bibinfo {volume} {130}},\ \bibinfo
  {pages} {154107} (\bibinfo {year} {2009})}\BibitemShut {NoStop}%
\bibitem [{\citenamefont {Bostr\"om}\ \emph {et~al.}(2013)\citenamefont
  {Bostr\"om}, \citenamefont {Veryazov}, \citenamefont {Aquilante},
  \citenamefont {Bondo~Pedersen},\ and\ \citenamefont
  {Lindh}}]{Bostrom:2014:321}%
  \BibitemOpen
  \bibfield  {author} {\bibinfo {author} {\bibfnamefont {J.}~\bibnamefont
  {Bostr\"om}}, \bibinfo {author} {\bibfnamefont {V.}~\bibnamefont {Veryazov}},
  \bibinfo {author} {\bibfnamefont {F.}~\bibnamefont {Aquilante}}, \bibinfo
  {author} {\bibfnamefont {T.}~\bibnamefont {Bondo~Pedersen}},\ and\ \bibinfo
  {author} {\bibfnamefont {R.}~\bibnamefont {Lindh}},\ }\bibfield  {title}
  {\bibinfo {title} {Analytical gradients of the second-order m{\o}ller-plesset
  energy using cholesky decompositions},\ }\href
  {https://doi.org/10.1002/qua.24563} {\bibfield  {journal} {\bibinfo
  {journal} {Int. J. Quantum Chem.}\ }\textbf {\bibinfo {volume} {114}},\
  \bibinfo {pages} {321} (\bibinfo {year} {2013})}\BibitemShut {NoStop}%
\bibitem [{\citenamefont {Thouless}(1960)}]{THOULESS1960225}%
  \BibitemOpen
  \bibfield  {author} {\bibinfo {author} {\bibfnamefont {D.}~\bibnamefont
  {Thouless}},\ }\bibfield  {title} {\bibinfo {title} {Stability conditions and
  nuclear rotations in the hartree-fock theory},\ }\href
  {https://doi.org/10.1016/0029-5582(60)90048-1} {\bibfield  {journal}
  {\bibinfo  {journal} {Nucl. Phys.}\ }\textbf {\bibinfo {volume} {21}},\
  \bibinfo {pages} {225} (\bibinfo {year} {1960})}\BibitemShut {NoStop}%
\bibitem [{\citenamefont {Wecker}\ \emph {et~al.}(2015)\citenamefont {Wecker},
  \citenamefont {Hastings}, \citenamefont {Wiebe}, \citenamefont {Clark},
  \citenamefont {Nayak},\ and\ \citenamefont {Troyer}}]{PhysRevA.92.062318}%
  \BibitemOpen
  \bibfield  {author} {\bibinfo {author} {\bibfnamefont {D.}~\bibnamefont
  {Wecker}}, \bibinfo {author} {\bibfnamefont {M.~B.}\ \bibnamefont
  {Hastings}}, \bibinfo {author} {\bibfnamefont {N.}~\bibnamefont {Wiebe}},
  \bibinfo {author} {\bibfnamefont {B.~K.}\ \bibnamefont {Clark}}, \bibinfo
  {author} {\bibfnamefont {C.}~\bibnamefont {Nayak}},\ and\ \bibinfo {author}
  {\bibfnamefont {M.}~\bibnamefont {Troyer}},\ }\bibfield  {title} {\bibinfo
  {title} {Solving strongly correlated electron models on a quantum computer},\
  }\href {https://doi.org/10.1103/physreva.92.062318} {\bibfield  {journal}
  {\bibinfo  {journal} {Phys. Rev. A}\ }\textbf {\bibinfo {volume} {92}},\
  \bibinfo {pages} {062318} (\bibinfo {year} {2015})}\BibitemShut {NoStop}%
\bibitem [{\citenamefont {Reck}\ \emph {et~al.}(1994)\citenamefont {Reck},
  \citenamefont {Zeilinger}, \citenamefont {Bernstein},\ and\ \citenamefont
  {Bertani}}]{PhysRevLett.73.58}%
  \BibitemOpen
  \bibfield  {author} {\bibinfo {author} {\bibfnamefont {M.}~\bibnamefont
  {Reck}}, \bibinfo {author} {\bibfnamefont {A.}~\bibnamefont {Zeilinger}},
  \bibinfo {author} {\bibfnamefont {H.~J.}\ \bibnamefont {Bernstein}},\ and\
  \bibinfo {author} {\bibfnamefont {P.}~\bibnamefont {Bertani}},\ }\bibfield
  {title} {\bibinfo {title} {Experimental realization of any discrete unitary
  operator},\ }\href {https://doi.org/10.1103/physrevlett.73.58} {\bibfield
  {journal} {\bibinfo  {journal} {Phys. Rev. Lett.}\ }\textbf {\bibinfo
  {volume} {73}},\ \bibinfo {pages} {58} (\bibinfo {year} {1994})}\BibitemShut
  {NoStop}%
\bibitem [{\citenamefont {Seritan}\ \emph {et~al.}(2020)\citenamefont
  {Seritan}, \citenamefont {Bannwarth}, \citenamefont {Fales}, \citenamefont
  {Hohenstein}, \citenamefont {Isborn}, \citenamefont {Kokkila-Schumacher},
  \citenamefont {Li}, \citenamefont {Liu}, \citenamefont {Luehr}, \citenamefont
  {Snyder}, \citenamefont {Song}, \citenamefont {Titov}, \citenamefont
  {Ufimtsev}, \citenamefont {Wang},\ and\ \citenamefont
  {Mart{\'\i}nez}}]{Seritan:2021:e1494}%
  \BibitemOpen
  \bibfield  {author} {\bibinfo {author} {\bibfnamefont {S.}~\bibnamefont
  {Seritan}}, \bibinfo {author} {\bibfnamefont {C.}~\bibnamefont {Bannwarth}},
  \bibinfo {author} {\bibfnamefont {B.~S.}\ \bibnamefont {Fales}}, \bibinfo
  {author} {\bibfnamefont {E.~G.}\ \bibnamefont {Hohenstein}}, \bibinfo
  {author} {\bibfnamefont {C.~M.}\ \bibnamefont {Isborn}}, \bibinfo {author}
  {\bibfnamefont {S.~I.~L.}\ \bibnamefont {Kokkila-Schumacher}}, \bibinfo
  {author} {\bibfnamefont {X.}~\bibnamefont {Li}}, \bibinfo {author}
  {\bibfnamefont {F.}~\bibnamefont {Liu}}, \bibinfo {author} {\bibfnamefont
  {N.}~\bibnamefont {Luehr}}, \bibinfo {author} {\bibfnamefont {J.~W.}\
  \bibnamefont {Snyder}}, \bibinfo {author} {\bibfnamefont {C.}~\bibnamefont
  {Song}}, \bibinfo {author} {\bibfnamefont {A.~V.}\ \bibnamefont {Titov}},
  \bibinfo {author} {\bibfnamefont {I.~S.}\ \bibnamefont {Ufimtsev}}, \bibinfo
  {author} {\bibfnamefont {L.-P.}\ \bibnamefont {Wang}},\ and\ \bibinfo
  {author} {\bibfnamefont {T.~J.}\ \bibnamefont {Mart{\'\i}nez}},\ }\bibfield
  {title} {\bibinfo {title} {{TeraChem}: {A} graphical processing
  unit-accelerated electronic structure package for large-scale ab initio
  molecular dynamics},\ }\href {https://doi.org/10.1002/wcms.1494} {\bibfield
  {journal} {\bibinfo  {journal} {WIREs Comput. Mol. Sci.}\ }\textbf {\bibinfo
  {volume} {11}},\ \bibinfo {pages} {e1494} (\bibinfo {year}
  {2020})}\BibitemShut {NoStop}%
\bibitem [{\citenamefont {Eastman}\ \emph {et~al.}(2017)\citenamefont
  {Eastman}, \citenamefont {Swails}, \citenamefont {Chodera}, \citenamefont
  {McGibbon}, \citenamefont {Zhao}, \citenamefont {Beauchamp}, \citenamefont
  {Wang}, \citenamefont {Simmonett}, \citenamefont {Harrigan}, \citenamefont
  {Stern}, \citenamefont {Wiewiora}, \citenamefont {Brooks},\ and\
  \citenamefont {Pande}}]{OpenMM}%
  \BibitemOpen
  \bibfield  {author} {\bibinfo {author} {\bibfnamefont {P.}~\bibnamefont
  {Eastman}}, \bibinfo {author} {\bibfnamefont {J.}~\bibnamefont {Swails}},
  \bibinfo {author} {\bibfnamefont {J.~D.}\ \bibnamefont {Chodera}}, \bibinfo
  {author} {\bibfnamefont {R.~T.}\ \bibnamefont {McGibbon}}, \bibinfo {author}
  {\bibfnamefont {Y.}~\bibnamefont {Zhao}}, \bibinfo {author} {\bibfnamefont
  {K.~A.}\ \bibnamefont {Beauchamp}}, \bibinfo {author} {\bibfnamefont {L.-P.}\
  \bibnamefont {Wang}}, \bibinfo {author} {\bibfnamefont {A.~C.}\ \bibnamefont
  {Simmonett}}, \bibinfo {author} {\bibfnamefont {M.~P.}\ \bibnamefont
  {Harrigan}}, \bibinfo {author} {\bibfnamefont {C.~D.}\ \bibnamefont {Stern}},
  \bibinfo {author} {\bibfnamefont {R.~P.}\ \bibnamefont {Wiewiora}}, \bibinfo
  {author} {\bibfnamefont {B.~R.}\ \bibnamefont {Brooks}},\ and\ \bibinfo
  {author} {\bibfnamefont {V.~S.}\ \bibnamefont {Pande}},\ }\bibfield  {title}
  {\bibinfo {title} {{OpenMM} 7: {Rapid} development of high performance
  algorithms for molecular dynamics},\ }\href
  {https://doi.org/10.1371/journal.pcbi.1005659} {\bibfield  {journal}
  {\bibinfo  {journal} {PLOS Comput. Biol.}\ }\textbf {\bibinfo {volume}
  {13}},\ \bibinfo {pages} {e1005659} (\bibinfo {year} {2017})}\BibitemShut
  {NoStop}%
\bibitem [{\citenamefont {Anselmetti}\ \emph {et~al.}(2021)\citenamefont
  {Anselmetti}, \citenamefont {Wierichs}, \citenamefont {Gogolin},\ and\
  \citenamefont {Parrish}}]{Anselmetti2021}%
  \BibitemOpen
  \bibfield  {author} {\bibinfo {author} {\bibfnamefont {G.-L.~R.}\
  \bibnamefont {Anselmetti}}, \bibinfo {author} {\bibfnamefont
  {D.}~\bibnamefont {Wierichs}}, \bibinfo {author} {\bibfnamefont
  {C.}~\bibnamefont {Gogolin}},\ and\ \bibinfo {author} {\bibfnamefont {R.~M.}\
  \bibnamefont {Parrish}},\ }\bibfield  {title} {\bibinfo {title} {Local,
  expressive, quantum-number-preserving {VQE} ans\"atze for fermionic
  systems},\ }\href {https://doi.org/10.1088/1367-2630/ac2cb3} {\bibfield
  {journal} {\bibinfo  {journal} {New J. Phys.}\ }\textbf {\bibinfo {volume}
  {23}},\ \bibinfo {pages} {113010} (\bibinfo {year} {2021})}\BibitemShut
  {NoStop}%
\bibitem [{\citenamefont {O'Gorman}\ \emph {et~al.}(2019)\citenamefont
  {O'Gorman}, \citenamefont {Huggins}, \citenamefont {Rieffel},\ and\
  \citenamefont {Whaley}}]{OGorman:2019}%
  \BibitemOpen
  \bibfield  {author} {\bibinfo {author} {\bibfnamefont {B.}~\bibnamefont
  {O'Gorman}}, \bibinfo {author} {\bibfnamefont {W.~J.}\ \bibnamefont
  {Huggins}}, \bibinfo {author} {\bibfnamefont {E.~G.}\ \bibnamefont
  {Rieffel}},\ and\ \bibinfo {author} {\bibfnamefont {K.~B.}\ \bibnamefont
  {Whaley}},\ }\bibfield  {title} {\bibinfo {title} {Generalized swap networks
  for near-term quantum computing},\ }\bibfield  {journal} {\bibinfo  {journal}
  {arXiv preprint arXiv:1905.05118}\ }\href
  {https://doi.org/10.48550/arxiv.1905.05118} {10.48550/arxiv.1905.05118}
  (\bibinfo {year} {2019})\BibitemShut {NoStop}%
\bibitem [{\citenamefont {Lee}\ \emph {et~al.}(2018)\citenamefont {Lee},
  \citenamefont {Huggins}, \citenamefont {Head-Gordon},\ and\ \citenamefont
  {Whaley}}]{Lee:2019:311}%
  \BibitemOpen
  \bibfield  {author} {\bibinfo {author} {\bibfnamefont {J.}~\bibnamefont
  {Lee}}, \bibinfo {author} {\bibfnamefont {W.~J.}\ \bibnamefont {Huggins}},
  \bibinfo {author} {\bibfnamefont {M.}~\bibnamefont {Head-Gordon}},\ and\
  \bibinfo {author} {\bibfnamefont {K.~B.}\ \bibnamefont {Whaley}},\ }\bibfield
   {title} {\bibinfo {title} {Generalized unitary coupled cluster wave
  functions for quantum computation},\ }\href
  {https://doi.org/10.1021/acs.jctc.8b01004} {\bibfield  {journal} {\bibinfo
  {journal} {J. Chem. Theory Comput.}\ }\textbf {\bibinfo {volume} {15}},\
  \bibinfo {pages} {311} (\bibinfo {year} {2018})}\BibitemShut {NoStop}%
\bibitem [{\citenamefont {Gidopoulos}\ and\ \citenamefont
  {Theophilou}(1994)}]{Gidopoulos:1993:1067}%
  \BibitemOpen
  \bibfield  {author} {\bibinfo {author} {\bibfnamefont {N.}~\bibnamefont
  {Gidopoulos}}\ and\ \bibinfo {author} {\bibfnamefont {A.}~\bibnamefont
  {Theophilou}},\ }\bibfield  {title} {\bibinfo {title} {Hartree-fock equations
  determining the optimum set of spin orbitals for the expansion of excited
  states},\ }\href {https://doi.org/10.1080/01418639408240176} {\bibfield
  {journal} {\bibinfo  {journal} {Philos. Mag. B}\ }\textbf {\bibinfo {volume}
  {69}},\ \bibinfo {pages} {1067} (\bibinfo {year} {1994})}\BibitemShut
  {NoStop}%
\bibitem [{\citenamefont {Gidopoulos}\ \emph {et~al.}(2002)\citenamefont
  {Gidopoulos}, \citenamefont {Papaconstantinou},\ and\ \citenamefont
  {Gross}}]{Gidopoulos:2002:328}%
  \BibitemOpen
  \bibfield  {author} {\bibinfo {author} {\bibfnamefont {N.}~\bibnamefont
  {Gidopoulos}}, \bibinfo {author} {\bibfnamefont {P.}~\bibnamefont
  {Papaconstantinou}},\ and\ \bibinfo {author} {\bibfnamefont {E.}~\bibnamefont
  {Gross}},\ }\bibfield  {title} {\bibinfo {title}
  {Ensemble-\{Hartree{\textendash}Fock\} scheme for excited states. the
  optimized effective potential method},\ }\href
  {https://doi.org/10.1016/s0921-4526(02)00799-8} {\bibfield  {journal}
  {\bibinfo  {journal} {Physica B}\ }\textbf {\bibinfo {volume} {318}},\
  \bibinfo {pages} {328} (\bibinfo {year} {2002})}\BibitemShut {NoStop}%
\bibitem [{\citenamefont {Swope}\ \emph {et~al.}(1982)\citenamefont {Swope},
  \citenamefont {Andersen}, \citenamefont {Berens},\ and\ \citenamefont
  {Wilson}}]{Swope:1982:637}%
  \BibitemOpen
  \bibfield  {author} {\bibinfo {author} {\bibfnamefont {W.~C.}\ \bibnamefont
  {Swope}}, \bibinfo {author} {\bibfnamefont {H.~C.}\ \bibnamefont {Andersen}},
  \bibinfo {author} {\bibfnamefont {P.~H.}\ \bibnamefont {Berens}},\ and\
  \bibinfo {author} {\bibfnamefont {K.~R.}\ \bibnamefont {Wilson}},\ }\bibfield
   {title} {\bibinfo {title} {A computer simulation method for the calculation
  of equilibrium constants for the formation of physical clusters of molecules:
  {Application} to small water clusters},\ }\href
  {https://doi.org/10.1063/1.442716} {\bibfield  {journal} {\bibinfo  {journal}
  {J. Chem. Phys.}\ }\textbf {\bibinfo {volume} {76}},\ \bibinfo {pages} {637}
  (\bibinfo {year} {1982})}\BibitemShut {NoStop}%
\bibitem [{\citenamefont {Dubnikova}\ and\ \citenamefont
  {Lifshitz}(2000)}]{Dubnikova:2000:4489}%
  \BibitemOpen
  \bibfield  {author} {\bibinfo {author} {\bibfnamefont {F.}~\bibnamefont
  {Dubnikova}}\ and\ \bibinfo {author} {\bibfnamefont {A.}~\bibnamefont
  {Lifshitz}},\ }\bibfield  {title} {\bibinfo {title} {Isomerization of
  propylene oxide. quantum chemical calculations and kinetic modeling},\ }\href
  {https://doi.org/10.1021/jp994485t} {\bibfield  {journal} {\bibinfo
  {journal} {J. Phys. Chem. A}\ }\textbf {\bibinfo {volume} {104}},\ \bibinfo
  {pages} {4489} (\bibinfo {year} {2000})}\BibitemShut {NoStop}%
\bibitem [{\citenamefont {Crosson}\ and\ \citenamefont
  {Moffat}(2001)}]{Crosson:2001:2995}%
  \BibitemOpen
  \bibfield  {author} {\bibinfo {author} {\bibfnamefont {S.}~\bibnamefont
  {Crosson}}\ and\ \bibinfo {author} {\bibfnamefont {K.}~\bibnamefont
  {Moffat}},\ }\bibfield  {title} {\bibinfo {title} {Structure of a
  flavin-binding plant photoreceptor domain: {Insights} into light-mediated
  signal transduction},\ }\href {https://doi.org/10.1073/pnas.051520298}
  {\bibfield  {journal} {\bibinfo  {journal} {Proc. Natl. Acad. Sci.}\ }\textbf
  {\bibinfo {volume} {98}},\ \bibinfo {pages} {2995} (\bibinfo {year}
  {2001})}\BibitemShut {NoStop}%
\bibitem [{\citenamefont {Warshel}\ and\ \citenamefont
  {Levitt}(1976)}]{Warshel1976}%
  \BibitemOpen
  \bibfield  {author} {\bibinfo {author} {\bibfnamefont {A.}~\bibnamefont
  {Warshel}}\ and\ \bibinfo {author} {\bibfnamefont {M.}~\bibnamefont
  {Levitt}},\ }\bibfield  {title} {\bibinfo {title} {Theoretical studies of
  enzymic reactions: {Dielectric,} electrostatic and steric stabilization of
  the carbonium ion in the reaction of lysozyme},\ }\href
  {https://doi.org/10.1016/0022-2836(76)90311-9} {\bibfield  {journal}
  {\bibinfo  {journal} {J. Mol. Biol.}\ }\textbf {\bibinfo {volume} {103}},\
  \bibinfo {pages} {227} (\bibinfo {year} {1976})}\BibitemShut {NoStop}%
\bibitem [{\citenamefont {Hohenstein}\ \emph
  {et~al.}(2015{\natexlab{a}})\citenamefont {Hohenstein}, \citenamefont
  {Luehr}, \citenamefont {Ufimtsev},\ and\ \citenamefont
  {Mart{\'\i}nez}}]{Hohenstein:2015:224103}%
  \BibitemOpen
  \bibfield  {author} {\bibinfo {author} {\bibfnamefont {E.~G.}\ \bibnamefont
  {Hohenstein}}, \bibinfo {author} {\bibfnamefont {N.}~\bibnamefont {Luehr}},
  \bibinfo {author} {\bibfnamefont {I.~S.}\ \bibnamefont {Ufimtsev}},\ and\
  \bibinfo {author} {\bibfnamefont {T.~J.}\ \bibnamefont {Mart{\'\i}nez}},\
  }\bibfield  {title} {\bibinfo {title} {An atomic orbital-based formulation of
  the complete active space self-consistent field method on graphical
  processing units},\ }\href {https://doi.org/10.1063/1.4921956} {\bibfield
  {journal} {\bibinfo  {journal} {J. Chem. Phys.}\ }\textbf {\bibinfo {volume}
  {142}},\ \bibinfo {pages} {224103} (\bibinfo {year}
  {2015}{\natexlab{a}})}\BibitemShut {NoStop}%
\bibitem [{\citenamefont {Hohenstein}\ \emph
  {et~al.}(2015{\natexlab{b}})\citenamefont {Hohenstein}, \citenamefont
  {Bouduban}, \citenamefont {Song}, \citenamefont {Luehr}, \citenamefont
  {Ufimtsev},\ and\ \citenamefont {Mart{\'\i}nez}}]{Hohenstein:2015:014111}%
  \BibitemOpen
  \bibfield  {author} {\bibinfo {author} {\bibfnamefont {E.~G.}\ \bibnamefont
  {Hohenstein}}, \bibinfo {author} {\bibfnamefont {M.~E.~F.}\ \bibnamefont
  {Bouduban}}, \bibinfo {author} {\bibfnamefont {C.}~\bibnamefont {Song}},
  \bibinfo {author} {\bibfnamefont {N.}~\bibnamefont {Luehr}}, \bibinfo
  {author} {\bibfnamefont {I.~S.}\ \bibnamefont {Ufimtsev}},\ and\ \bibinfo
  {author} {\bibfnamefont {T.~J.}\ \bibnamefont {Mart{\'\i}nez}},\ }\bibfield
  {title} {\bibinfo {title} {Analytic first derivatives of floating occupation
  molecular orbital-complete active space configuration interaction on
  graphical processing units},\ }\href {https://doi.org/10.1063/1.4923259}
  {\bibfield  {journal} {\bibinfo  {journal} {J. Chem. Phys.}\ }\textbf
  {\bibinfo {volume} {143}},\ \bibinfo {pages} {014111} (\bibinfo {year}
  {2015}{\natexlab{b}})}\BibitemShut {NoStop}%
\bibitem [{\citenamefont {Mandal}\ \emph {et~al.}(2003)\citenamefont {Mandal},
  \citenamefont {Tahara},\ and\ \citenamefont {Meech}}]{Mandal:2004:1102}%
  \BibitemOpen
  \bibfield  {author} {\bibinfo {author} {\bibfnamefont {D.}~\bibnamefont
  {Mandal}}, \bibinfo {author} {\bibfnamefont {T.}~\bibnamefont {Tahara}},\
  and\ \bibinfo {author} {\bibfnamefont {S.~R.}\ \bibnamefont {Meech}},\
  }\bibfield  {title} {\bibinfo {title} {Excited-state dynamics in the green
  fluorescent protein chromophore},\ }\href {https://doi.org/10.1021/jp035816b}
  {\bibfield  {journal} {\bibinfo  {journal} {The Journal of Physical Chemistry
  B}\ }\textbf {\bibinfo {volume} {108}},\ \bibinfo {pages} {1102} (\bibinfo
  {year} {2003})}\BibitemShut {NoStop}%
\bibitem [{\citenamefont {Hohenstein}(2016)}]{Hohenstein:2016:174110}%
  \BibitemOpen
  \bibfield  {author} {\bibinfo {author} {\bibfnamefont {E.~G.}\ \bibnamefont
  {Hohenstein}},\ }\bibfield  {title} {\bibinfo {title} {Analytic formulation
  of derivative coupling vectors for complete active space configuration
  interaction wavefunctions with floating occupation molecular orbitals},\
  }\href {https://doi.org/10.1063/1.4966235} {\bibfield  {journal} {\bibinfo
  {journal} {J. Chem. Phys.}\ }\textbf {\bibinfo {volume} {145}},\ \bibinfo
  {pages} {174110} (\bibinfo {year} {2016})}\BibitemShut {NoStop}%
\bibitem [{\citenamefont {Parrish}\ \emph {et~al.}(2019)\citenamefont
  {Parrish}, \citenamefont {Hohenstein}, \citenamefont {McMahon},\ and\
  \citenamefont {Mart{\'\i}nez}}]{Parrish:2019:230401}%
  \BibitemOpen
  \bibfield  {author} {\bibinfo {author} {\bibfnamefont {R.~M.}\ \bibnamefont
  {Parrish}}, \bibinfo {author} {\bibfnamefont {E.~G.}\ \bibnamefont
  {Hohenstein}}, \bibinfo {author} {\bibfnamefont {P.~L.}\ \bibnamefont
  {McMahon}},\ and\ \bibinfo {author} {\bibfnamefont {T.~J.}\ \bibnamefont
  {Mart{\'\i}nez}},\ }\bibfield  {title} {\bibinfo {title} {Quantum computation
  of electronic transitions using a variational quantum eigensolver},\ }\href
  {https://doi.org/10.1103/physrevlett.122.230401} {\bibfield  {journal}
  {\bibinfo  {journal} {Phys. Rev. Lett.}\ }\textbf {\bibinfo {volume} {122}},\
  \bibinfo {pages} {230401} (\bibinfo {year} {2019})}\BibitemShut {NoStop}%
\bibitem [{\citenamefont {Parrish}\ \emph {et~al.}(2021)\citenamefont
  {Parrish}, \citenamefont {Anselmetti},\ and\ \citenamefont
  {Gogolin}}]{Parrish2021mcvqegradient}%
  \BibitemOpen
  \bibfield  {author} {\bibinfo {author} {\bibfnamefont {R.~M.}\ \bibnamefont
  {Parrish}}, \bibinfo {author} {\bibfnamefont {G.-L.~R.}\ \bibnamefont
  {Anselmetti}},\ and\ \bibinfo {author} {\bibfnamefont {C.}~\bibnamefont
  {Gogolin}},\ }\bibfield  {title} {\bibinfo {title} {Analytical ground- and
  excited-state gradients for molecular electronic structure theory from hybrid
  {Quantum/Classical} methods}\ }\href
  {https://doi.org/10.48550/arxiv.2110.05040} {10.48550/arxiv.2110.05040}
  (\bibinfo {year} {2021})\BibitemShut {NoStop}%
\bibitem [{\citenamefont {Clements}\ \emph {et~al.}(2016)\citenamefont
  {Clements}, \citenamefont {Humphreys}, \citenamefont {Metcalf}, \citenamefont
  {Kolthammer},\ and\ \citenamefont {Walmsley}}]{clements2016optimal}%
  \BibitemOpen
  \bibfield  {author} {\bibinfo {author} {\bibfnamefont {W.~R.}\ \bibnamefont
  {Clements}}, \bibinfo {author} {\bibfnamefont {P.~C.}\ \bibnamefont
  {Humphreys}}, \bibinfo {author} {\bibfnamefont {B.~J.}\ \bibnamefont
  {Metcalf}}, \bibinfo {author} {\bibfnamefont {W.~S.}\ \bibnamefont
  {Kolthammer}},\ and\ \bibinfo {author} {\bibfnamefont {I.~A.}\ \bibnamefont
  {Walmsley}},\ }\bibfield  {title} {\bibinfo {title} {Optimal design for
  universal multiport interferometers},\ }\href@noop {} {\bibfield  {journal}
  {\bibinfo  {journal} {Optica}\ }\textbf {\bibinfo {volume} {3}},\ \bibinfo
  {pages} {1460} (\bibinfo {year} {2016})}\BibitemShut {NoStop}%
\bibitem [{\citenamefont {Li}\ \emph {et~al.}(2017)\citenamefont {Li},
  \citenamefont {Yang}, \citenamefont {Peng},\ and\ \citenamefont
  {Sun}}]{PhysRevLett.118.150503}%
  \BibitemOpen
  \bibfield  {author} {\bibinfo {author} {\bibfnamefont {J.}~\bibnamefont
  {Li}}, \bibinfo {author} {\bibfnamefont {X.}~\bibnamefont {Yang}}, \bibinfo
  {author} {\bibfnamefont {X.}~\bibnamefont {Peng}},\ and\ \bibinfo {author}
  {\bibfnamefont {C.-P.}\ \bibnamefont {Sun}},\ }\bibfield  {title} {\bibinfo
  {title} {Hybrid quantum-classical approach to quantum optimal control},\
  }\href {https://doi.org/10.1103/physrevlett.118.150503} {\bibfield  {journal}
  {\bibinfo  {journal} {Phys. Rev. Lett.}\ }\textbf {\bibinfo {volume} {118}},\
  \bibinfo {pages} {150503} (\bibinfo {year} {2017})}\BibitemShut {NoStop}%
\bibitem [{\citenamefont {Mitarai}\ \emph {et~al.}(2018)\citenamefont
  {Mitarai}, \citenamefont {Negoro}, \citenamefont {Kitagawa},\ and\
  \citenamefont {Fujii}}]{PhysRevA.98.032309}%
  \BibitemOpen
  \bibfield  {author} {\bibinfo {author} {\bibfnamefont {K.}~\bibnamefont
  {Mitarai}}, \bibinfo {author} {\bibfnamefont {M.}~\bibnamefont {Negoro}},
  \bibinfo {author} {\bibfnamefont {M.}~\bibnamefont {Kitagawa}},\ and\
  \bibinfo {author} {\bibfnamefont {K.}~\bibnamefont {Fujii}},\ }\bibfield
  {title} {\bibinfo {title} {Quantum circuit learning},\ }\href
  {https://doi.org/10.1103/physreva.98.032309} {\bibfield  {journal} {\bibinfo
  {journal} {Phys. Rev. A}\ }\textbf {\bibinfo {volume} {98}},\ \bibinfo
  {pages} {032309} (\bibinfo {year} {2018})}\BibitemShut {NoStop}%
\bibitem [{\citenamefont {Bergholm}\ \emph {et~al.}(2018)\citenamefont
  {Bergholm}, \citenamefont {Izaac}, \citenamefont {Schuld}, \citenamefont
  {Gogolin}, \citenamefont {Alam}, \citenamefont {Ahmed}, \citenamefont
  {Arrazola}, \citenamefont {Blank}, \citenamefont {Delgado}, \citenamefont
  {Jahangiri} \emph {et~al.}}]{PennyLane}%
  \BibitemOpen
  \bibfield  {author} {\bibinfo {author} {\bibfnamefont {V.}~\bibnamefont
  {Bergholm}}, \bibinfo {author} {\bibfnamefont {J.}~\bibnamefont {Izaac}},
  \bibinfo {author} {\bibfnamefont {M.}~\bibnamefont {Schuld}}, \bibinfo
  {author} {\bibfnamefont {C.}~\bibnamefont {Gogolin}}, \bibinfo {author}
  {\bibfnamefont {M.~S.}\ \bibnamefont {Alam}}, \bibinfo {author}
  {\bibfnamefont {S.}~\bibnamefont {Ahmed}}, \bibinfo {author} {\bibfnamefont
  {J.~M.}\ \bibnamefont {Arrazola}}, \bibinfo {author} {\bibfnamefont
  {C.}~\bibnamefont {Blank}}, \bibinfo {author} {\bibfnamefont
  {A.}~\bibnamefont {Delgado}}, \bibinfo {author} {\bibfnamefont
  {S.}~\bibnamefont {Jahangiri}}, \emph {et~al.},\ }\bibfield  {title}
  {\bibinfo {title} {Pennylane: {Automatic} differentiation of hybrid
  quantum-classical computations},\ }\bibfield  {journal} {\bibinfo  {journal}
  {arXiv preprint arXiv:1811.04968}\ }\href
  {https://doi.org/10.48550/arxiv.1811.04968} {10.48550/arxiv.1811.04968}
  (\bibinfo {year} {2018})\BibitemShut {NoStop}%
\bibitem [{\citenamefont {Mari}\ \emph {et~al.}(2021)\citenamefont {Mari},
  \citenamefont {Bromley},\ and\ \citenamefont {Killoran}}]{Mari2021}%
  \BibitemOpen
  \bibfield  {author} {\bibinfo {author} {\bibfnamefont {A.}~\bibnamefont
  {Mari}}, \bibinfo {author} {\bibfnamefont {T.~R.}\ \bibnamefont {Bromley}},\
  and\ \bibinfo {author} {\bibfnamefont {N.}~\bibnamefont {Killoran}},\
  }\bibfield  {title} {\bibinfo {title} {Estimating the gradient and
  higher-order derivatives on quantum hardware},\ }\bibfield  {journal}
  {\bibinfo  {journal} {Phys. Rev. A}\ }\textbf {\bibinfo {volume} {103}},\
  \href {https://doi.org/10.1103/physreva.103.012405}
  {10.1103/physreva.103.012405} (\bibinfo {year} {2021})\BibitemShut {NoStop}%
\bibitem [{\citenamefont {Wierichs}\ \emph {et~al.}(2021)\citenamefont
  {Wierichs}, \citenamefont {Izaac}, \citenamefont {Wang},\ and\ \citenamefont
  {Lin}}]{wierichs2021general}%
  \BibitemOpen
  \bibfield  {author} {\bibinfo {author} {\bibfnamefont {D.}~\bibnamefont
  {Wierichs}}, \bibinfo {author} {\bibfnamefont {J.}~\bibnamefont {Izaac}},
  \bibinfo {author} {\bibfnamefont {C.}~\bibnamefont {Wang}},\ and\ \bibinfo
  {author} {\bibfnamefont {C.~Y.-Y.}\ \bibnamefont {Lin}},\ }\href
  {https://doi.org/10.22331/q-2022-03-30-677} {\bibinfo {title} {General
  parameter-shift rules for quantum gradients}} (\bibinfo {year} {2021}),\
  \Eprint {https://arxiv.org/abs/2107.12390} {arXiv:2107.12390 [quant-ph]}
  \BibitemShut {NoStop}%
\end{thebibliography}
\end{document}